\documentclass[review]{elsarticle}
\pdfoutput=1
\usepackage{lineno,hyperref}
\usepackage{amsmath,amssymb,amsfonts}
\usepackage[T1]{fontenc}
\usepackage[normalem]{ulem}
\modulolinenumbers[5]

\usepackage{comment}

\usepackage{amsmath, upgreek}

\usepackage{caption}
\usepackage{subcaption}
\usepackage[section]{placeins}
\usepackage{multirow} 
\usepackage{bm}
\usepackage{xcolor}
\usepackage{multirow}

\makeatletter
\def\ps@pprintTitle{%
 \let\@oddhead\@empty
 \let\@evenhead\@empty
 \def\@oddfoot{\centerline{\thepage}}%
 \let\@evenfoot\@oddfoot}
\makeatother

\biboptions{authoryear}

\begin{document}

\begin{frontmatter}

\title{A novel \textit{in vivo} approach to assess strains of the human abdominal wall
under known intraabdominal pressure}

\tnotetext[mytitlenote]{This is the accepted version of a paper published  in the Journal of the Mechanical Behavior of Biomedical Materials, 125 (2022) 104902. The final publication is available at https://doi.org/10.1016/j.jmbbm.2021.104902}


\author[wilis]{Izabela Lubowiecka}

\author[wilis]{Katarzyna Szepietowska\corref{mycorrespondingauthor}}
\cortext[mycorrespondingauthor]{Corresponding author}
\ead{katszepi@pg.edu.pl}

\author[wilis]{Agnieszka Tomaszewska}
\author[oio]{Pawe\l{} Micha\l{} Bielski}
\author[gumed]{Micha\l{} Chmielewski}
\author[gumed]{Monika Lichodziejewska-Niemierko}
\author[oio]{Czes\l{}aw Szymczak}
\address[wilis]{Faculty of Civil and Environmental Engineering, Gda\'nsk University of Technology, Gda\'nsk, Poland}
\address[oio]{Faculty of Mechanical Engineering and Ship Technology, Gda\'nsk University of Technology, Gda\'nsk, Poland}
\address[gumed]{Department of Nephrology, Transplantology and Internal Medicine, Medical University of Gda\'nsk, Gda\'nsk, Poland}

\begin{abstract}

The study concerns mechanical behaviour of a living human abdominal wall. A better mechanical understanding of a human abdominal wall and recognition of its material properties is required to find  mechanically compatible surgical meshes to significantly improve the treatment of ventral hernias. 

A non-invasive methodology, based on \textit{in vivo} optical measurements is proposed to determine strains of abdominal wall corresponding to a known intraabdominal pressure. 
The measurement is performed in the course of a standard procedure of peritoneal dialysis. A dedicated experimental stand is designed for the experiment.
The photogrammetric technique is employed to recover the three-dimensional surface geometry of  the anterior abdominal wall at the initial and terminal instants of the dialysis. This corresponds to two deformation states, before and after filling the abdominal cavity with dialysis fluid. 

The study provides information on strain fields of living human abdominal wall. The inquiry is aimed at principal strains and their directions, observed at the level from -10\% to 17\%. 
The intraabdominal pressure related to the amount of introduced dialysis fluid  measured within the medical procedure covers the range 11-18.5 cmH$_2$O. 

The methodology leads to the deformation state of the abdominal wall according to the corresponding loading conditions. Therefore, the study is  a step towards an identification of mechanical properties of living human abdominal wall.

\end{abstract}


\end{frontmatter}

\linenumbers

\section{Introduction}

The paper addresses mechanics of human anterior abdominal wall.  The main motivation to the study comes from the need of the ventral hernia treatment improvement. Mechanical compatibility of the implanted surgical mesh with the abdominal wall is assured in order to achieve successful hernia repair \citep{junge2001elasticity}. Thus the mechanical behaviour of  surgical meshes  and abdominal tissues has been studied and reported in literature \cite[see extensive review of ][]{Deeken2017}. Compatibility between an implant and an abdominal wall can be assessed experimentally  {\citep{junge2001elasticity,anurov2012biomechanical}} according to various  criteria  \citep{maurer2014mechanical} or by computer simulations \citep{simon2016prostheses,todros2018computational,he2020numerical}. The computational approach allows for  \textit{in silico} testing of various cases and  empowers  optimisation of ventral hernia treatment \citep{szymczak2017two}. To accurately predict performance of the system composed of abdominal wall and implant, the mathematical model requires relevant constitutive modelling of both implant and abdominal wall.

  Mechanics of the separate components of human and animal abdominal wall were studied in the literature. The connective tissues of the abdominal wall  were investigated  \citep{kirilova2011experimental,astruc2018characterization}, since they are considered decisive in the context of hernia. The literature  features experimental studies on linea alba \citep{cooney2016uniaxial,levillain2016contribution} including also a choice of an accurate material model  \citep{santamaria2015material}. Mechanics of abdominal wall muscles was also investigated, in terms of both passive  \citep{calvo2014determination,hernandez2011mechanical}  and active  behaviour \citep{grasa2016active}.
  
  \cite{podwojewski2014mechanical}  investigated  strains in an entire abdominal wall (with all layers) subjected to  pressure load in the following states: intact, incised and repaired one with implanted surgical mesh.  \cite{tran2014contribution} used a similar setup to investigate the contribution of abdominal layers to the response of the abdominal wall. Rat abdominal wall parameters were also identified by an inflation  test   \citep{mahalingam2017burst}. \cite{LERUYET2020103683} compared different suturing techniques studying strain field  on external surface of the myofascial abdominal wall in five \textit{post mortem} human specimens subjected to pressure with the use of digital image correlation method.

    These studies on the mechanical properties of the abdominal wall were conducted on specimens \textit{post mortem},  which can be considered a limitation. The mechanical performance of abdominal wall observed in such tests may not fully correspond to a living tissue performence under physiological conditions. \textit{Post mortem} investigation has certain limitations \citep{Deeken2017}, such as: effects of freezing, dehydration, \textit{rigor mortis} and the  analysis often based on aged donors, which may not correspond to the response of younger tissues. The existing numerical models of abdominal wall \citep{hernandez2013understanding,pachera2016numerical,tuset2019implementation} are  mainly based on material properties identified from \textit{ex vivo} test, to possibly interfere the prediction accuracy of a living human abdominal wall performance. While mechanical properties of living tissues are highly variable within testing, the material parameters uncertainty  should be also  taken into account. \cite{cmbbe2020} showed the uncertainty impact of material model parameters on the abdominal wall model response. To reduce the uncertainty of the model outcome, mechanics of  human  abdominal wall should be widely investigated  \textit{in vivo},  methodologies leading to personalised models should be developed. 
  
  Few existing \textit{in vivo} studies on abdominal wall can be divided into two main groups: the first  employing medical imaging  and the second based on measurements of displacements on the external surface of the abdomen. \cite{tran2016abdominal}  employed shear wave elastography and local stiffness measurements to asses elasticity of a human abdominal wall.  \cite{linek2019supersonic}  focused on shear wave elastography of lateral muscles,  controlled muscle activity using electromyography.  \cite{jourdan2021semiautomatic} studied deformation of abdominal wall during breathing using dynamic MRI image registration.
  
 Displacements measurements of an external abdomen surface   were employed to study strains of human abdominal wall in selected activities  \citep{szymczak2012investigation}. Full field measurement with the use of digital image correlation method \citep{breier2017evaluation} was employed to assess  strains on the external surface of an abdomen during different movements.  \cite{todros20193d} involved laser scanning to study deformation during abdominal muscle contractions.

  \cite{song2006elasticity,song2006mechanical}  identified Young's modulus of living human abdominal wall based on displacements of markers on the skin within  changes of intraabdominal pressure during laparoscopic repair. Similar study was performed to study children's abdominal wall \citep{zhou2020abdominal}. Nevertheless, a credible computational model of abdominal wall may require employing hyperelastic constitutive law including anisotropy related to fiber architecture \cite[like e.g. the model of][]{tuset2019implementation}.  \cite{simon2015developing} developed the idea of measurements carried out during laparoscopic repair on an animal model  to identify parameters of a hyperelastic isotropic material law  and their  spatial distribution in the rabbit abdominal wall \citep{simon2017towards}, additionally, to study the effects of implanting surgical meshes on an animal model \citep{simon2018mechanical}. The \textit{in vivo} performance of a surgical mesh implanted in abdominal wall became another research issue.   \cite{kahan2018combined} proposed a methodology to study \textit{in vivo}  strains of implanted surgical meshes on an animal model with the use of fluoroscopic images. The elongation of surgical meshes during side bending of torso was studied \textit{in vivo} by \cite{lubowiecka2020vivo}. However, the parameters of  hyperelastic material law or the spatially distributed material parameters of the living human abdominal wall are still missing.
  
    The study is aimed at a methodology for  deformations assessment of the living human abdominal wall subjected to known  intraabdominal pressures, which can be used for further identification of the material model of the abdominal wall \cite[see extensive overview of identification methods by][]{avril2008overview}.  In the course of  peritoneal dialysis it is possible to measure intraperitoneal pressure in a non-invasive manner  \citep{Durand1996}. Following \cite{intraperitoneal_Al-Hwiesh2011}, the intraperitoneal pressure should be routinely measured within the peritoneal dialysis. Moreover,  no statistical difference occurs between both pressures, intraperitoneal and intraabdominal, in either  erect or supine positions.
   Peritoneal dialysis (PD) is a type of renal-replacement therapy applied for patients with end-stage renal disease. The method is based on insertion of a given volume of dialysis fluid into the abdominal cavity. Patient's uremic toxins and excess water are being removed from the organism to the dialysis fluid through the processes of diffusion and osmosis.      As the patients have a peritoneal catheter implanted, it is possible to measure
 intraperitoneal pressure when the abdominal cavity is newly filled with dialysis fluid \citep{PerezDiaz2017}. 
   It opens the door to perform a deformation measurement of abdominal wall while  introduction of the dialysis fluid when intraabdominal pressure can  be  measured too.    Photogrammetric method is employed to assess the displacements of abdominal wall during  pressure change by extracting 3D geometry from 2D photos taken in different experimental stages. A special experimental hospital-oriented stand was designed for that purpose.   A 3-D reconstruction of the geometry is based on a set of images taken at various angles,  hence the measurements cost is relatively low,  compared to commercial Digital Image Correlation systems or laser scanning. Although photogrammetry has been mainly developed as a tool in geodesy to measure large-scale object, and applied to engineering structures, see e.g. \citep{armesto2009fem}, it has also been used in biomechanics, e.g. to capture deformation of surgical meshes \cite{barone2015impact}. The preliminary results of the measurements of the abdominal wall displacements \textit{in vivo} during PD were presented in the conference paper  \citep{lubowiecka2018ssta}. In the  study, the methodology is  developed, the experimental stand is proposed, last but not  least, the  obtained principal strains  of the human living abdominal wall are presented and discussed.

\section{Materials and Methods}
\subsection{Description of patients}

Seven patients suffering from end-stage kidney disease,  regularly subjected to PD have been included in the study. Obesity acts as an excluding criterion since a thick layer of subcutaneous fat makes the results hard to interpret in terms of concluding on the behaviour of abdominal wall main components based on strains measured on its external layer. The parameters of the patients and major health condition data of their anterior abdominal walls are included in Table \ref{Table_patient}.

\begin{table}[ht]
\begin{tabular}{p{0.5cm}p{0.5cm}p{0.5cm}p{0.8cm}p{0.8cm}p{0.9cm}p{6cm}}
\hline
No & sex & age & height [m]  &weight [kg] & BMI [kg/m$^2$] &issues with   abdominal wall\slash hernia \\

\hline
P1          & F            & 46           & 1.64             & 70  & 26.0   & --- \\
P2           & F            & 55           & 1.63             & 62    &23.2    & suspicion of a   hernia   \\
P3           & M            & 65           & 1.73             & 82& 27.2       &supra-umbilical hernia, diastasis recti \\
P4          & M            & 64           & 1.82             & 74      &22.3       & two   hernias \\
P5           & M            & 34           & 1.74             & 81     &26.8     & umbilical hernia  \\
P6         & M            & 34           & 1.83             & 67   &20.0      &  --- \\
P7         & M            & 47           & 1.76             & 82    &26.5   & repair of   right inguinal hernia with Lichtenstein method and synthetic implant\\
\hline
\end{tabular}
\caption{Characteristics of the patients}\label{Table_patient}
\end{table}

\subsection{Description of experiments}

The measurements are fully non-invasive performed during  PD procedure, regular for the patients. All participants submitted a consent to participate in the study,  the study protocol was approved by the local Ethics Committee (NKBBN 314/2018).

\begin{figure}
\centering

\begin{subfigure}[b]{0.40\textwidth} 
\includegraphics[width=\textwidth]{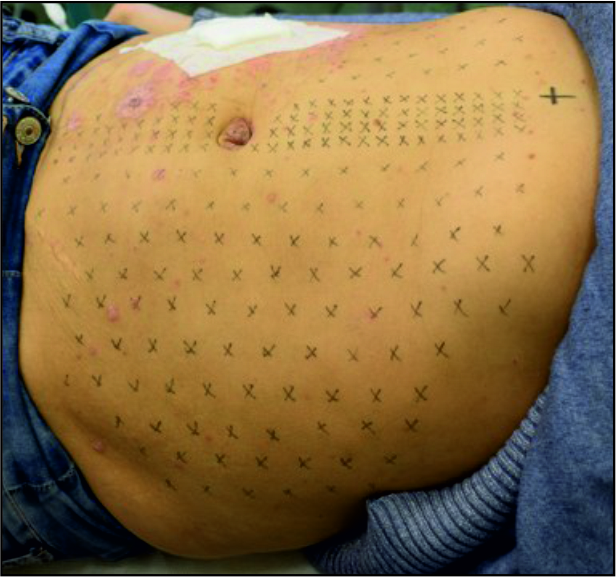}
\caption{Reference stage - empty abdominal cavity}\label{fig_etap_drained}
\end{subfigure}
 \begin{subfigure}[b]{0.40\textwidth} 
\includegraphics[width=\textwidth]{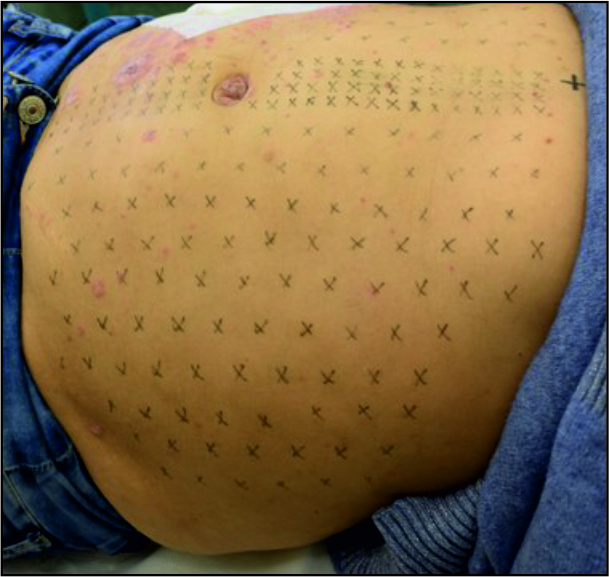}
\caption{Deformed stage - abdominal cavity filled with dialysis fluid}\label{fig_etap_filled}
\end{subfigure}
    \caption{Stages of data acquisition. Reference and deformed stages of an abdominal wall}
    \label{fig_etapy}
\end{figure}

 The patients were subjected to  optical measurements  of their abdominal wall geometry in two stages of the PD procedure. 
 In both stages the abdomen is in different geometric state.  In the first stage the abdominal cavity is drained  from dialysis fluid (Figure \ref{fig_etap_drained}), it serves as a reference state for the proposed analysis. In the second stage (Figure \ref{fig_etap_filled}), the abdominal cavity is filled with a new portion of two-liter dialysis fluid \cite[see][]{Durand1996}. The intraabdominal pressure grows in the filling course, the abdominal wall deforms. The post-filling instant marks a deformed state of the abdominal wall, regarded in the analysis. Hence it is possible to register geometry change  related to the change of  intraabdominal pressure acting on an abdominal wall. Regular pattern-forming marks are made on the investigated abdominal wall applied for triangulation and 3D reconstruction of the abdominal wall required to assess its deformation. In both states the 3D reconstruction of the abdomen geometry is determined with the help of photogrammetry. 
 
\begin{figure}[ht]\centering
    \includegraphics[width=\textwidth]{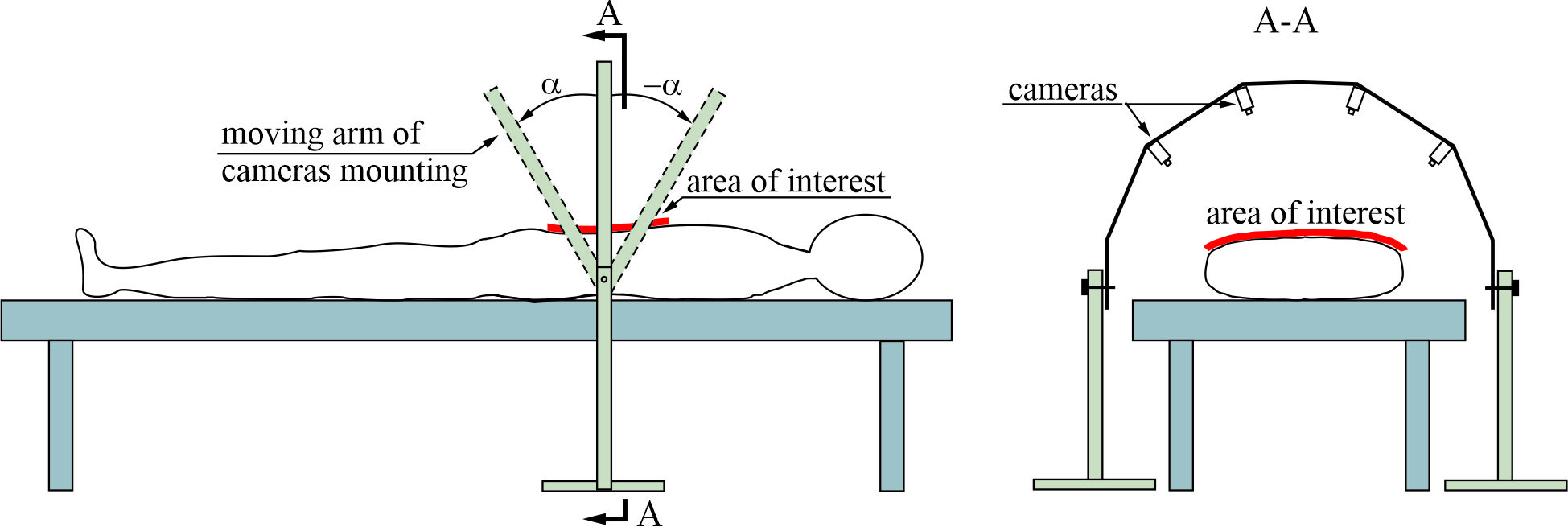}
    \caption{Experimental stand}
    \label{fig_exp}
\end{figure}

\begin{figure}[ht]\centering
    \includegraphics[width=0.5\textwidth]{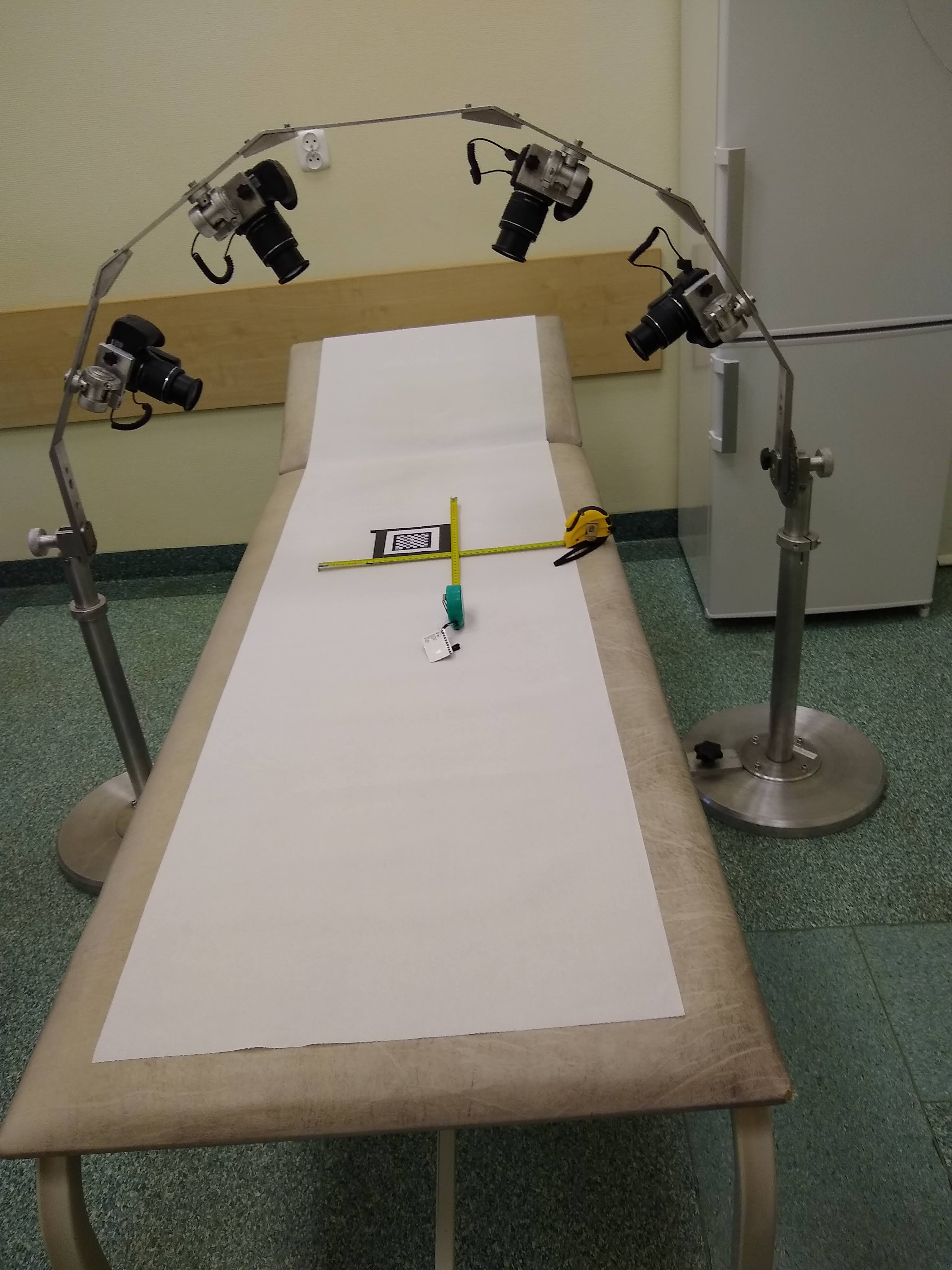}
    \caption{Experimental stand prepared for testing}   \label{fig_stand_zdj}
\end{figure}

The 3D measurements are taken from 2D data based on photographs of the subject. The photos are taken simultaneously by four cameras installed in a specially designed stand (patent application number P.438555). The experimental set-up is shown in Figure  \ref{fig_exp}, and the photo of the apparatus is in Figure  \ref{fig_stand_zdj}. The stand is situated above the area of interest (abdominal wall). The distance between the cameras and the abdomen is approximately 0.5 m. The camera mounting arm is ready to rotate about the horizontal line situated perpendicularly to cranio-caudal axis of the patient. The pictures were taken in several positions of the arm changing the angle $\alpha$ every {15}$^{\circ}$. That enabled to collect a relevant picture set of the entire area of interest in a given state. The photos are bound to overlap, to reconstruct the 3D geometry of the abdominal wall. The Nikon D3300 cameras with 24 Mpix matrix, equipped with Nikon AF-S DX VR Nikkor 18-55mm f/3.5-5.6G II lenses have been used. Simultaneous photos are taken with the use of self-timers installed in each camera, launched in a contactless mode by an additional timer operated by a researcher. Data acquisition is presented in Figure \ref{fig_data_aq}. Calibration is provided to properly scale the collected photos. Before a proper measurement, a chequerboard calibration plate  was photographed to capture the relative positions of mounted cameras. Later in the postprocessing the data, these relative
positions were used to scale and reference the point cloud. Breathing influence is reduced while the pictures are shot in a full exhalation of a patient.

The abdominal pressure is measured  when the abdominal cavity is filled with a new portion of the dialysis fluid. The pressure is measured with the use of manometer connected to the dialysis bag, as regarded by \cite{PerezDiaz2017}.

\subsection{Post-processing and calculation}

\subsubsection{ Calculation of strains based on 3D geometry reconstruction}

 Image processing was executed with the help of Agisoft Metashape software, a commercial photogrammetry package, and RawTherapee, an open-source image-editing software.

During the camera alignment phase, relative positions of cameras are calculated and sparse point cloud of the measured object is created. Because of the subject motion (mainly respiration), photos taken from different angles in different time captured slightly different geometry. It resulted in some difficulties with automatic camera alignment.  To enhance the process, we manually marked several characteristic points in each image and linked these points with neighbouring images. Camera alignment  was then optimised to ensure that these points overlap in respective projection planes. Full resolution photos have been used to perform camera alignment and build the initial sparse point cloud. 
Coordinates of a specific camera rig position were exported. These coordinates were used later in photogrammetric models of measured subjects, so that each state (empty and full) of a subject was represented in the same coordinate system and the same scale.

\begin{figure}[tbh]\centering
\begin{subfigure}[b]{0.4\textwidth} 
\includegraphics[width=\textwidth]{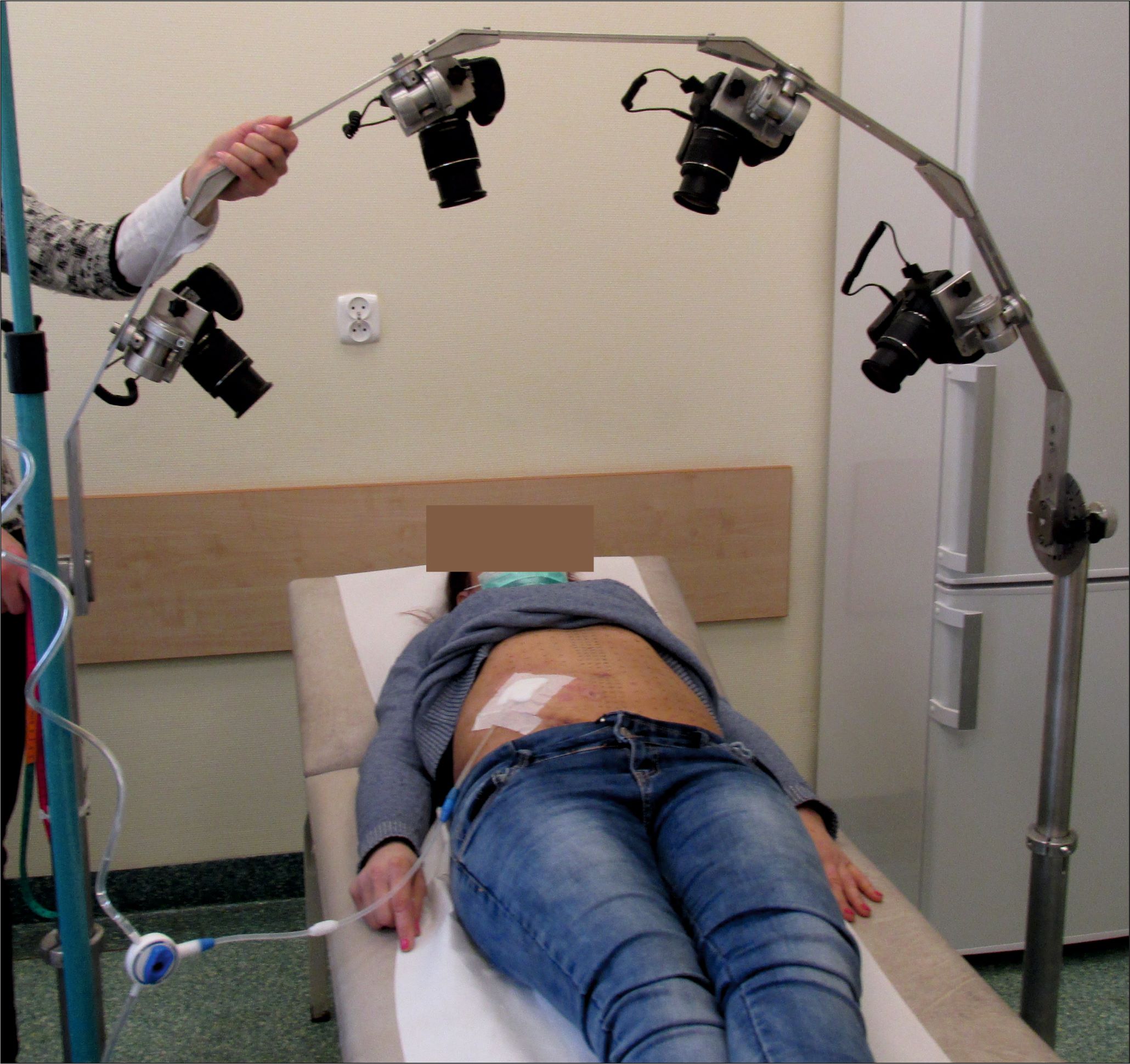}
\caption{Data acquisition}\label{fig_data_aq}
\end{subfigure}
\begin{subfigure}[b]{0.4\textwidth} 
\includegraphics[width=\textwidth]{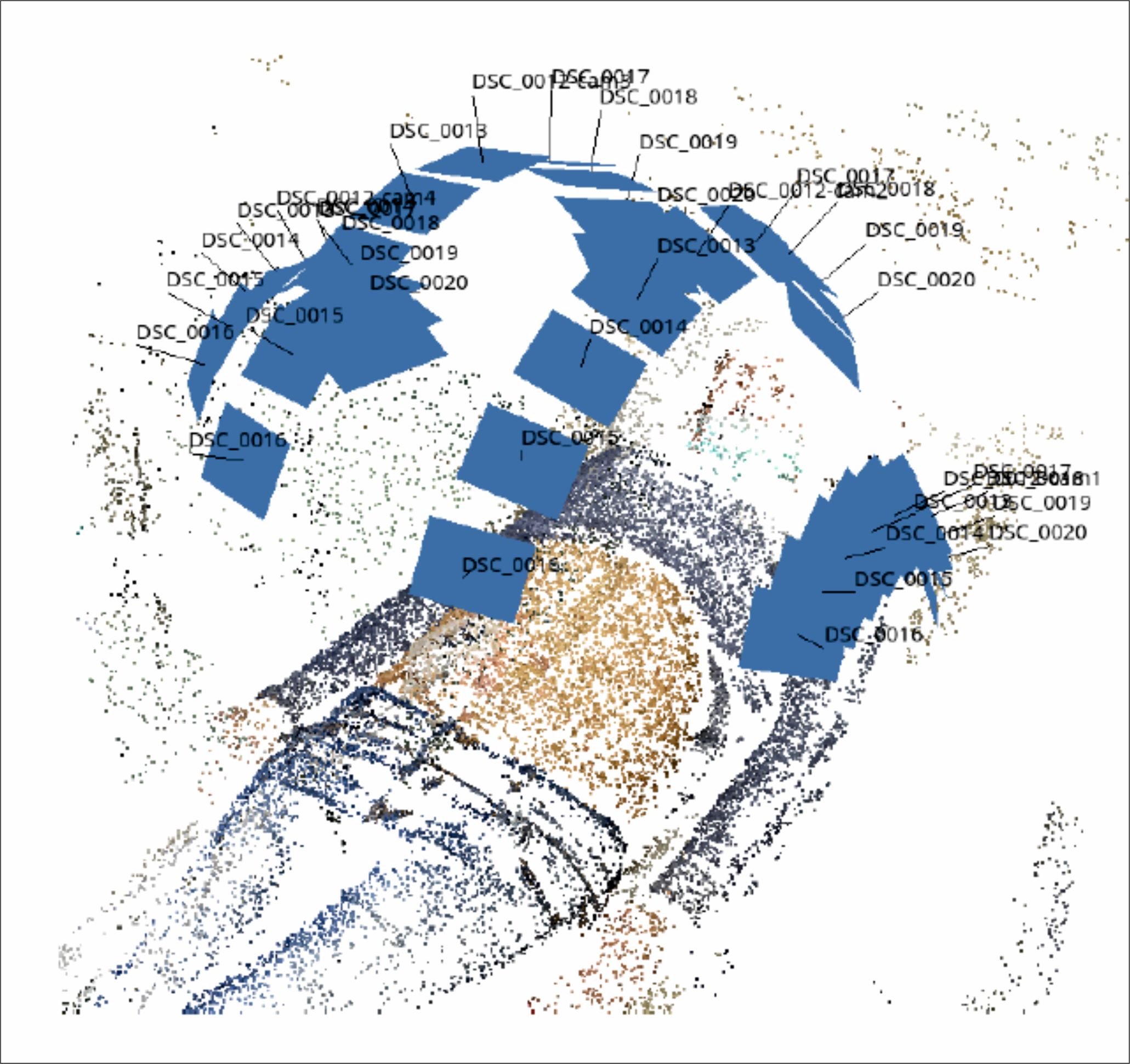}
\caption{  Point cloud and camera positions } \label{fig_cloud_point}
\end{subfigure}
    \caption{Photogrammetric measurements and corresponding first stage of data postprocessing}
    \label{fig_cloud}
\end{figure}

\begin{figure}[tbh]\centering

    \includegraphics[width=0.7\textwidth]{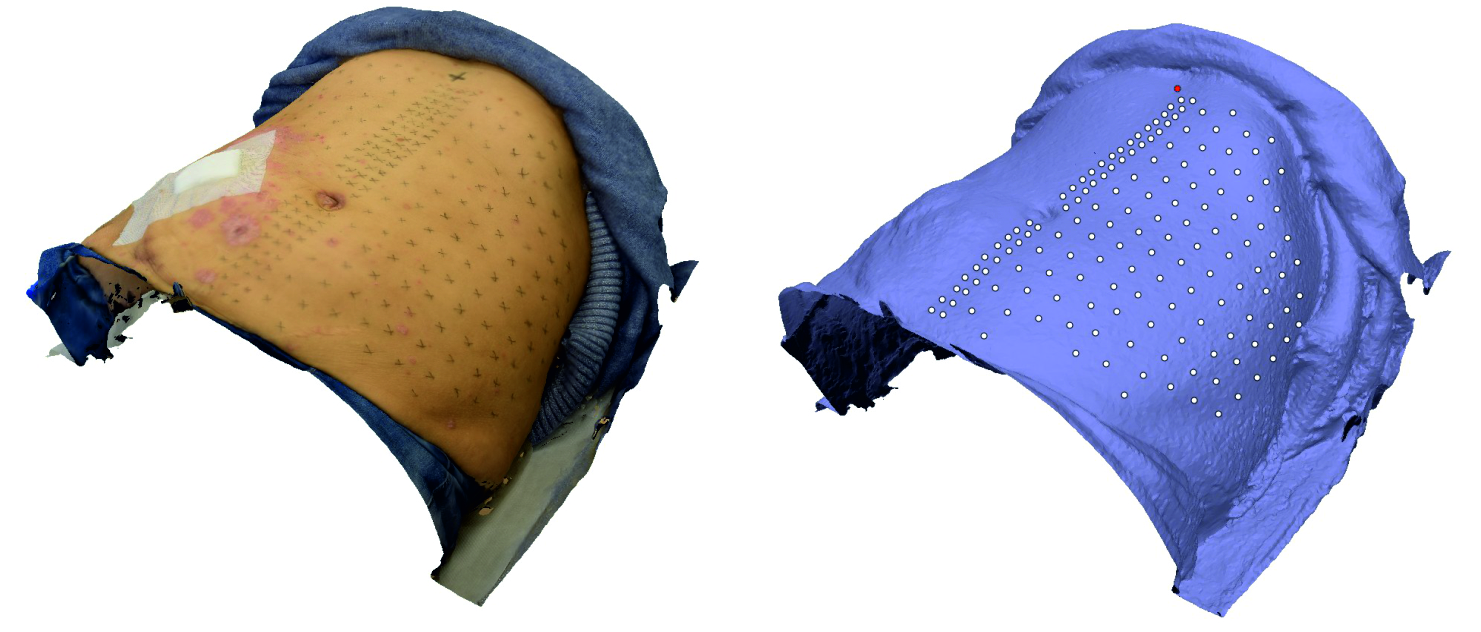}
    \caption{Photogrammetric data postprocessing - 3D reconstruction of abdominal wall geometry}
    \label{fig_data_post}
    
\end{figure}

Based on the collected photos, the point cloud was generated,  see Figure \ref{fig_cloud_point},  the mapped positions of the four cameras during data acquisition are also shown. The process was conducted in both reference and deformed states.
Next, the abdominal wall geometry was reconstructed in a three-dimensional space, as shown in Figure \ref{fig_data_post}.

The point cloud of abdominal wall was meshed using the 2D Delaunay triangulation. Twin meshes were obtained for both regarded states, as the same points were used in triangulation.

Local coordinate systems were defined in each triangle of the mesh, making two in-plane axes, x y, and a single normal axis  created in accordance with the vertices. To calculate nodal displacements and strains, each triangle is considered to follow the plane strain state.  

The strains are calculated using linear and nonlinear approach \citep{Fung2001,Holzapfel2000,wriggers2008nonlinear}. Engineering (Cauchy) strains and Green-Lagrange strain measure are computed to compare the two approaches which are both present in the literature about the mechanics of the abdominal wall tissues and implants.

In the Lagrange description the motion of a particle in time  can be described by $\mathbf{x}=\mathbf{x}(\mathbf{X},t)$, where  $\mathbf{X}$ denotes its position in reference and $\mathbf{x}$ in deformed configurations. In this description, the displacement field is described by $\mathbf{U}$,
\begin{equation}
 \mathbf{U}(\mathbf{X},t)=\mathbf{x}(\mathbf{X},t)-\mathbf{X}.   
\end{equation}
It can be also expressed in the Euler description as
\begin{equation}
 \mathbf{u}(\mathbf{x},t)=\mathbf{x}-\mathbf{X}(\mathbf{x},t).   
\end{equation}
The deformation gradient is defined as
\begin{equation}
\mathbf{F}=\frac{\partial \mathbf{x}}{\partial\mathbf{X}}, \quad
    F_{i,I}=\frac{\partial x_i}{\partial X_I}
\end{equation}
and the Green-Lagrange strain tensor $\mathbf{E}$ takes the following form 
\begin{equation}
    \mathbf{E}=\frac{1}{2}\left( \mathbf{F}^\top \mathbf{F} - \mathbf{I}\right).
\end{equation}
The strain tensor can be also expressed in terms of displacement gradient  as
\begin{equation}
    E_{ij}=\frac{1}{2}\Big[ \frac{\partial U_j}{\partial X_i}+\frac{\partial U_i}{\partial X_j}+\frac{\partial U_\alpha}{\partial X_i}\frac{\partial U_\alpha}{\partial X_j}\Big]
\end{equation}
that in the Euler description Euler-Almansi strain tensor takes the form of
\begin{equation}
    e_{ij}=\frac{1}{2}\Big[ \frac{\partial u_j}{\partial x_i}+\frac{\partial u_i}{\partial x_j}-\frac{\partial u_\alpha}{\partial x_i}\frac{\partial u_\alpha}{\partial x_j}\Big]. \label{eq:euler_strain}
\end{equation}
The deformation gradient is computed following nonlinear finite element solution for a triangular element, where  $\mathbf{X}$ and  $\mathbf{x}$ are interpolated by shape functions $N_I$
\begin{equation}
    \mathbf{X_e}=\sum_{I=1}^n N_I(\bm{\xi})\mathbf{X}_I, \quad 
    \mathbf{x_e}=\sum_{I=1}^n N_I(\bm{\xi})\mathbf{x}_I,
\end{equation}
where $\bm{\xi}$ are the coordinates of a reference element \citep{wriggers2008nonlinear}. Deformation gradient of a finite element $\mathbf{F}_e$ can be defined as:
\begin{equation}
   \mathbf{F}_e=\mathbf{j}_e \mathbf{J}_e^{-1},
\end{equation}
where
\begin{equation}
    \mathbf{j}_e=\frac{\partial \mathbf{x}}{\partial \bm{\xi}},  \quad \mathbf{J}_e=\frac{\partial \mathbf{X}}{\partial \bm{\xi}}, \label{eq_jacobi}
\end{equation}

The triangular element with shape functions
\begin{equation}
    N_1=1-\xi-\upeta, \quad N_2=\xi, \quad N_3=\upeta,
\end{equation}
leads to the Jacobi matrix (\ref{eq_jacobi}) in the following form:
\begin{equation}
 \mathbf{J}_e=   \begin{bmatrix}
    X_l-X_k & X_m-X_k\\
    Y_l-Y_k & Y_m-Y_k
    \end{bmatrix}
\end{equation}
and
\begin{equation}
 \mathbf{j}_e=   \begin{bmatrix}
    x_l-x_k & x_m-x_k\\
    y_l-y_k & y_m-y_k
    \end{bmatrix}.
\end{equation}

When components of displacement $u_i$ are small, $e_{ij}$ (\ref{eq:euler_strain}) can be reduced to Cauchy infinitesimal strain tensor:
\begin{equation}
\varepsilon_{ij}=\displaystyle\frac{1}{2}\Big[\displaystyle\frac{\partial u_j}{\partial x_i}+\frac{\partial u_i}{\partial x_j}\Big],
\label{strain_tensor}
\end{equation}
its unabridged form reads (\ref{strains_xy}) 

\begin{equation}
\varepsilon_{xx}=\frac{\partial u}{\partial x},\quad \varepsilon_{yy}=\frac{\partial v}{\partial y},\quad \varepsilon_{xy}=\frac{1}{2}\left(\frac{\partial u}{\partial y}+\frac{\partial v}{\partial x} \right)=\varepsilon_{yx}, 
\label{strains_xy}
\end{equation}
where $(x,y)$ are local nodal coordinates of every element, $(u$,$v)$ are the corresponding  nodal displacements. In this case   distinction between Lagrange and Euler description disappear.
The strain tensor is defined in every triangle. The strains are assumed constant in element in both,  reference and  deformed configurations, thus nodal displacement are considered only. 
Two principal strains $\varepsilon_1$ and $ \varepsilon_2$ satisfy   the equation (\ref{principal_strain}), which describes eigenvalue problem, the same for Green-Largange and Cauchy strain tensor.

\begin{equation}
\vert \varepsilon_{ij}-\varepsilon \delta_{ij} \vert = 0
\label{principal_strain}
\end{equation}

Each principal strain is associated with its principal axis, whose direction cosines $v^{(1)}_{j}$, $v^{(2)}_{j}$ fulfil the following:

\begin{equation}
(\varepsilon_{ij}-\varepsilon_1 \delta_{ij}) v^{(1)}_{j} = 0,\quad (i = 1,2).
\label{principal_strain2}
\end{equation}

\begin{figure}[tbh]\centering
    \includegraphics[width=0.4\textwidth]{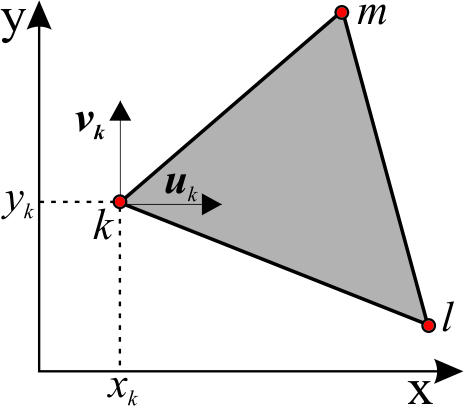}
    \caption{Triangular element in plane strain state}
    \label{triangle-workspace}
\end{figure}

In a triangular element (Figure \ref{triangle-workspace}) undergoing plane strain,  displacements of the  nodes of the element are defined by two linear polynomials \citep{zienkiewicz2005finite}.

\begin{equation}
\begin{aligned}
     u = \alpha_1 + \alpha_2x + \alpha_3y, \\
     v = \alpha_4 + \alpha_5x + \alpha_6y.
\end{aligned}
\end{equation}

A set of six constants $\alpha_k$ can be obtained by deriving two sets of three equations:
\begin{equation}
     u_k = \alpha_1 + \alpha_2x_k + \alpha_3y_k, \\
     u_l = \alpha_1 + \alpha_2x_l + \alpha_3y_l, \\
     u_m = \alpha_1 + \alpha_2x_m + \alpha_3y_m, \\
\end{equation}
with respect to $\alpha_1$, $\alpha_2$, $\alpha_3$, affected by on $u_k$, $u_l$, $u_m$, eventually  finding the displacement
\begin{equation}
     u = \frac{1}{2\Delta}[(a_k+b_kx+c_ky)u_i+(a_l+b_lx+c_ly)u_j+(a_m+b_mx+c_my)u_m],
\end{equation}
where $\Delta$ is the area of the triangle $i$, $j$, $m$ is a determinant form:

\begin{equation}
    \Delta = \frac{1}{2} \det \begin{bmatrix}
    1 & x_k & y_k \\
    1 & x_l & y_l \\
    1 & x_m & y_m
    \end{bmatrix}
\end{equation}
and 
\begin{equation}
     a_k= x_ly_m-x_my_l,\quad b_k=y_l-y_m=y_{lm},\quad c_k= x_m-x_l=x_{ml}.
\end{equation}
Other coefficients $a$, $b$, $c$ are derived  the same way, calculating $k$, $l$, $m$. 

The displacement $v$ can be obtained  similar by
\begin{equation}
     v = \frac{1}{2\Delta}[(a_k+b_kx+c_ky)v_k+(a_l+b_lx+c_ly)v_l+(a_m+b_mx+c_my)v_m],
\end{equation}

The displacement functions defined this way assure continuity on the borders with neighbouring elements, since they are linear along each triangle side. Thus the same displacements of nodes assure the same displacements between the triangles. The components $\varepsilon_x$, $\varepsilon_y$, $\gamma_{xy}$ of plane strain tensor read:  

\begin{equation}
    \begin{Bmatrix}
    \varepsilon_x \\ \varepsilon_y \\ 2\varepsilon_{xy} \end{Bmatrix} =
    \dfrac{1}{2\Delta} \begin{bmatrix}
    b_k & 0 & b_l & 0 & b_m & 0 \\
    0 & c_k & 0 & c_l & 0 & c_m \\
    c_k & b_1 & c_l & b_m & c_m & b_m \end{bmatrix}
    \begin{Bmatrix}
    u_k \\ v_k \\ u_l \\ v_l \\ u_m \\ v_m \end{Bmatrix}.
\end{equation}

The measured abdominal wall of every tested patient was reconstructed in both states, reference and deformed. The displacements of every node and strains of each triangular element were computed as described above. All calculations were performed in MATLAB environment.

\subsubsection{Bayesian analysis}

The presented results can be combined with the information on strain of abdominal wall available in literature  to find  their probabilistic distribution parameters.  Following \cite{Straub2015}, the classical Bayesian updating (statistical inference) can be applied to find probabilistic distribution of the model parameters based on measurements. The  conditional probability of $\theta$ given the observations $\mathbf{d}$  can be calculated according to Bayes' rule as 
\begin{equation}
f_{\theta \mid \mathbf{d}}(\theta)=aL(\theta)f_{\theta}(\theta),
\end{equation}
where $f_{\theta}$ is the prior probability density function (PDF) updated to the posterior PDF $f_{\theta \mid \mathbf{d}}$ with the use of the observations $\mathbf{d}$ represented by  the likelihood function $L(\theta)$ and the normalising constant $a$ is
\begin{equation}
    a=\frac{1}{\int_{\mathrm{R}^n} L(\theta)f_\theta(\theta)d\theta}.
\end{equation}

The  analysis is focused on maximum $\varepsilon_1$ observed in abdominal wall of a patient because maximum strains can be interesting from the viewpoint of choice of surgical mesh with an appropriate  strain range. Let us assume that $\mathcal{X}$, standing for the maximum $\varepsilon_1$, is a Gaussian variable with unknown mean $\theta$ and fixed standard deviation. The likelihood function $L(\theta)$ is a Gaussian variable with mean $\theta$ and fixed standard deviation $\sigma_\mathcal{X}=3.3\%$ with a conjugate prior --- Gaussian variable whose parameters are $\mu_\theta=13.25\%$ and $\sigma_\theta=5.27\%$. The parameters of prior distributions are assumed that 90th percentile is 20\%, (in top range of values reported in the literature for similar value of intraabdominal pressure, \cite{LERUYET2020103683})   and 1st percentile is 1\%. In this conjugate prior case the analytical solution is known and the posterior  $\theta$ is Gaussian with the mean following the formula:

\begin{equation}
    \mu_{\theta \mid \mathbf{d}}=\frac{\mu_\theta / \sigma_\theta ^2 + m \bar{d}/\sigma_\mathcal{X}^2}{1/\sigma_\theta ^2+m/\sigma_\mathcal{X}^2},
\end{equation}
and the standard deviation
\begin{equation}
        \sigma_{\theta \mid \mathbf{d}}=\left(\frac{1}{1/\sigma_\theta ^2+m/\sigma_\mathcal{X}^2} \right)^{1/2},
\end{equation}
where $m=7$ is the number of observations $\mathbf{d}$ (Table \ref{Table_pressure}) of the random variable $\mathcal{X}$ and $\bar{d}$ is the mean of $\mathbf{d}$. Since $\theta$ is a Gaussian variable,  the predictive distribution of $\mathcal{X}$ with unknown mean  may be found, it is also a Gaussian variable  with the same mean, 
 $\mu=\mu_{\theta \mid \mathbf{d}}$ and standard deviation, $\sigma=\sqrt{\sigma_{\theta \mid \mathbf{d}}^2+\sigma_\mathcal{X}^2 }$.

\section{Results and discussion}

\begin{figure}[tbh]\centering
\begin{subfigure}[b]{\textwidth}  
    \includegraphics[width=\textwidth]{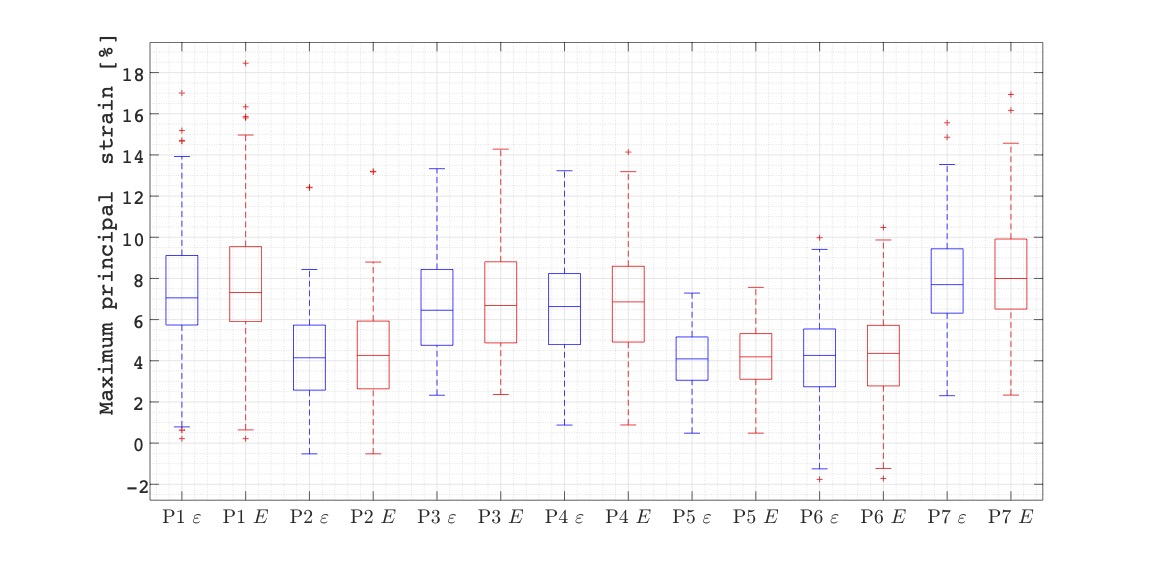}
    \caption{Maximum principal strain}
    \label{boxplot_max}
\end{subfigure}

\begin{subfigure}[b]{\textwidth}  
    \includegraphics[width=\textwidth]{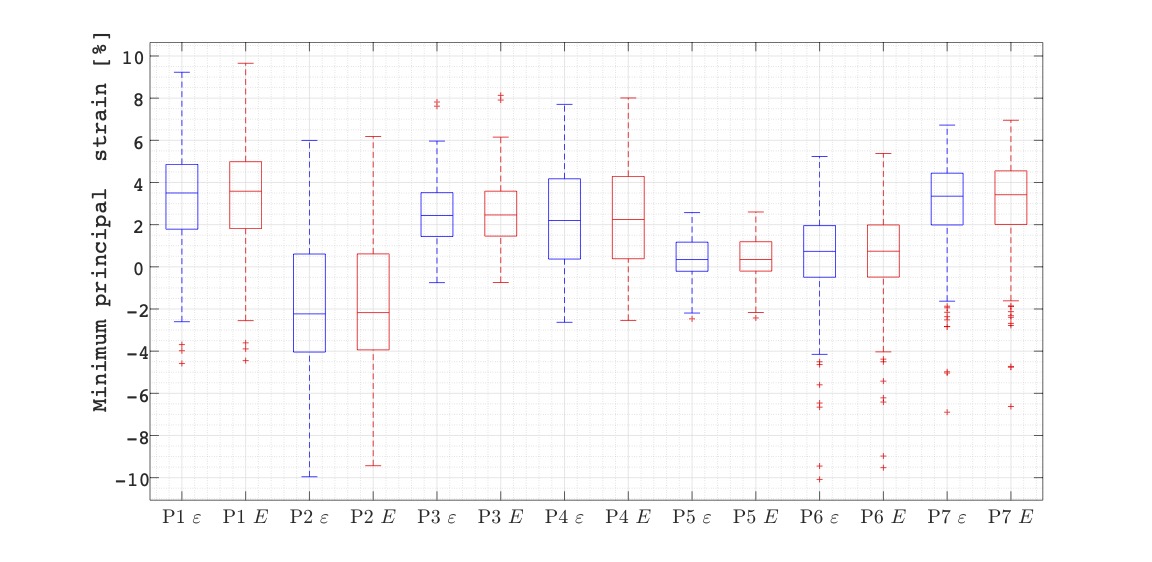}
    \caption{Minimum principal strain}
    \label{boxplot_min}
    \end{subfigure}
    \caption{Boxplots of principal  Cauchy ($\varepsilon$, blue boxes) and Green-Lagrange ($E$, red boxes) strains. The central line is the median, edges of box are the 25th and 75th percentiles (distance between top and bottom edge is interquartile range), whiskers extend to furthest data points but outliers are marked by red '+' (beyond 1.5 times interquartile range away from top or bottom of the box)} \label{boxplot}
\end{figure}

The intraabdominal pressure measured after full introduction of dialysis fluid  is presented in Table \ref{Table_pressure}. The table  also includes the minimum,  maximum and median values of principal strains, both Cauchy (engineering) and Green-Lagrange, observed in all cases. Maximum absolute difference between Green-Lagrange and engineering strains equals 1.45\% (7.9\% relative difference with respect to Green-Lagrange value). This value was observed in case of P1.  Mean absolute difference, in terms of patients, between Green-Lagrange and engineering strains equals  0.89\% (6.2\% relative difference with respect to Green-Lagrange value) in case of  maximum value of $\varepsilon_1$, 0.23\%  (3.7\% relative difference with respect to Green-Lagrange value) in case of minimum value of $\varepsilon_2$ and 0.19\% (3.1\% relative difference with respect to Green-Lagrange value) and 0.04\% (1.9\% relative difference with respect to Green-Lagrange value) for median of  $\varepsilon_1$ and $\varepsilon_2$, respectively.  The differences between engineering and Green-Lagrange strains statistics can be  observed on the boxplots  (Figure \ref{boxplot}). The  further discussion is  based on the values of Cauchy strains, since the analysis of both, Cauchy and Green-Lagrange strain values are relatively similar and lead to similar conclusions.
 
\begin{figure}[tbh]\centering
    \includegraphics[width=0.8\textwidth]{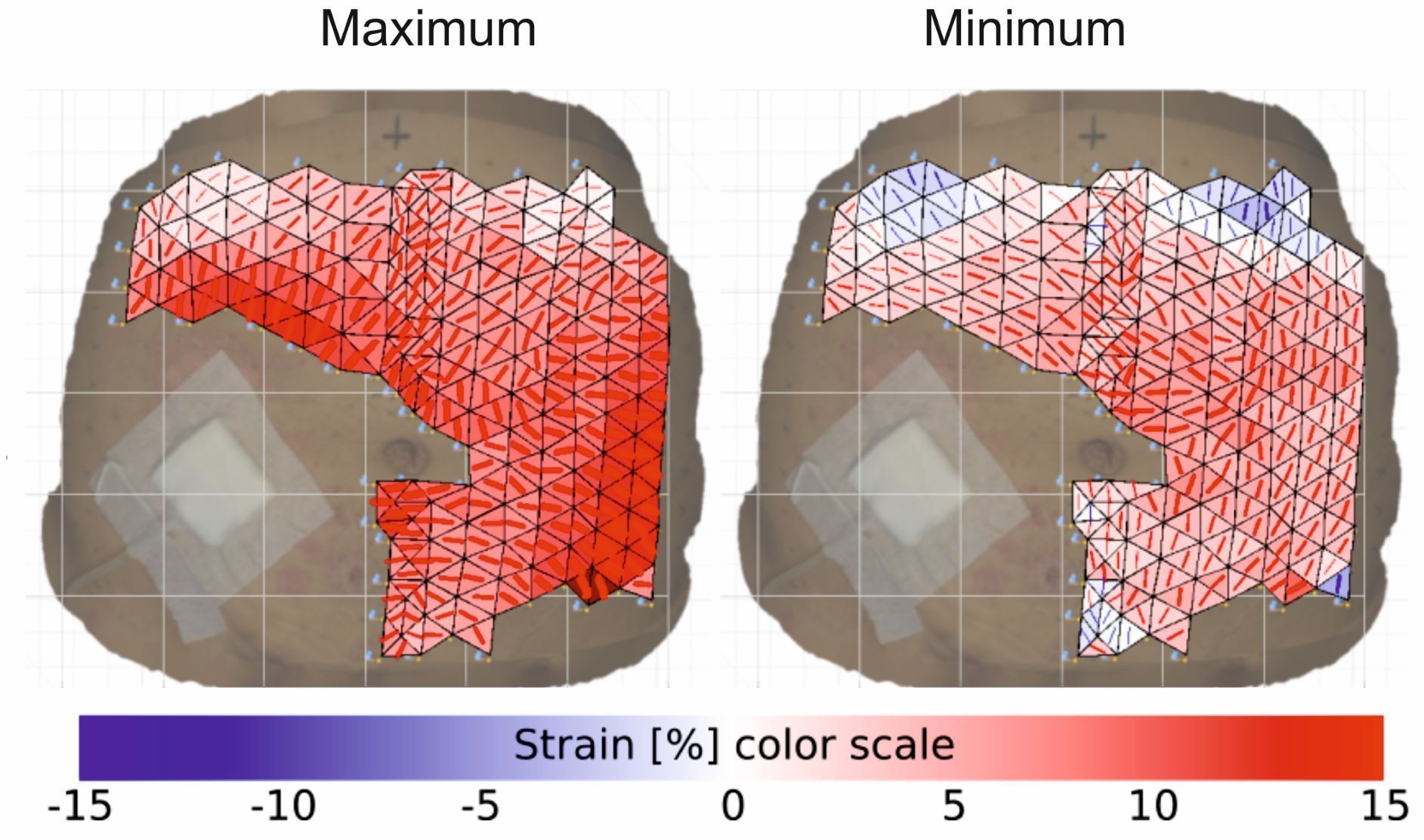}
    \caption{Principal maximum and minimum  Cauchy strains with principal directions for Patient P1 }
    \label{fig:sub05}
\end{figure}

\begin{figure}[tbh]\centering
    \includegraphics[width=0.8\textwidth]{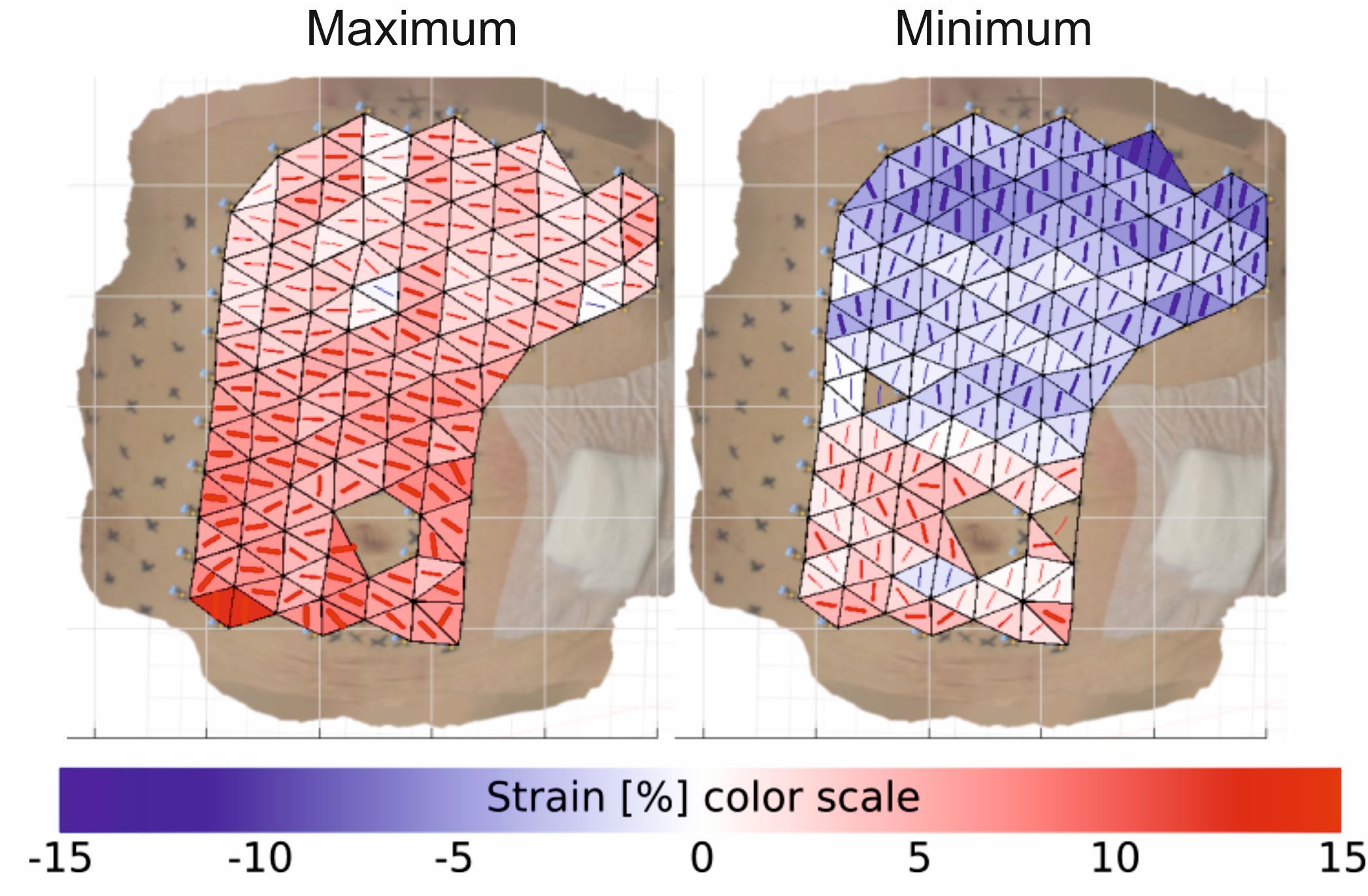}
    \caption{Principal maximum and minimum  Cauchy strains with principal directions for Patient P2}
    \label{fig:maps_sub06}
\end{figure}

\begin{figure}[tbh]\centering
    \includegraphics[width=0.8\textwidth]{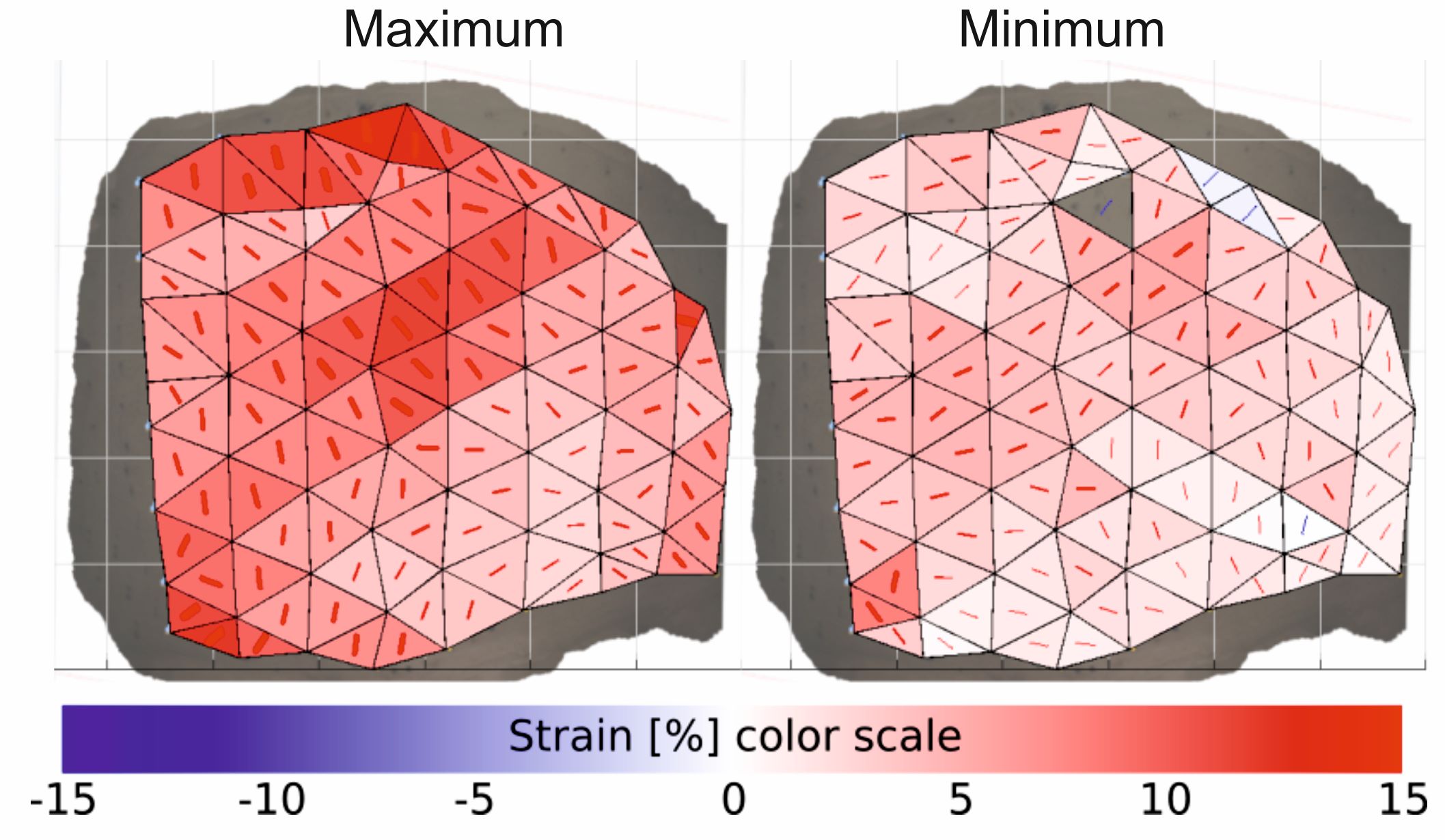}
    \caption{Principal maximum and minimum Cauchy  strains with principal directions for Patient P3}
    \label{fig:maps_sub07}
\end{figure}

\begin{figure}[tbh]\centering
    \includegraphics[width=0.8\textwidth]{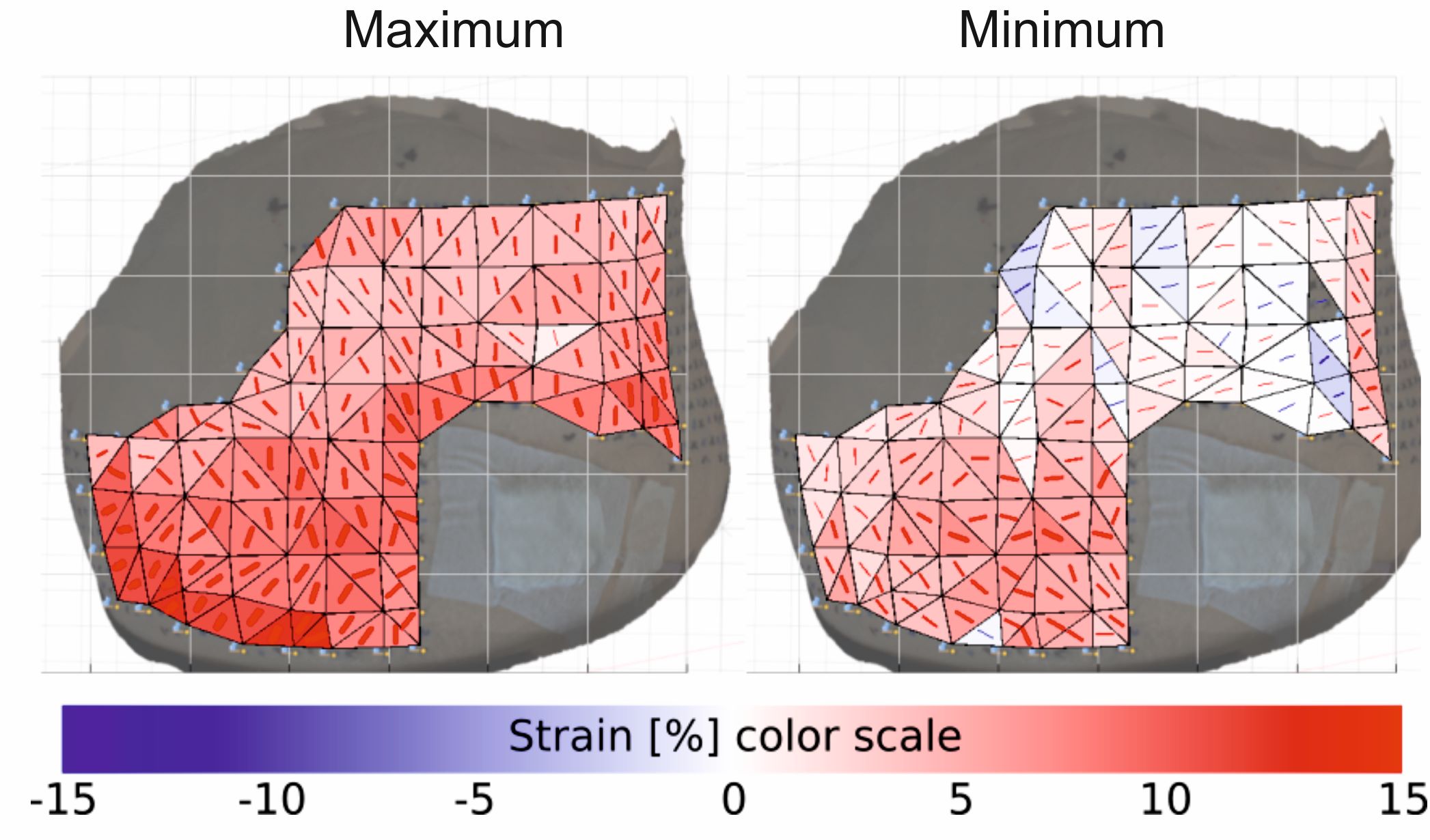}
    \caption{Principal maximum and minimum Cauchy strains with principal directions for Patient P4}
    \label{fig:maps_sub08}
\end{figure}

\begin{figure}[tbh]\centering
    \includegraphics[width=0.8\textwidth]{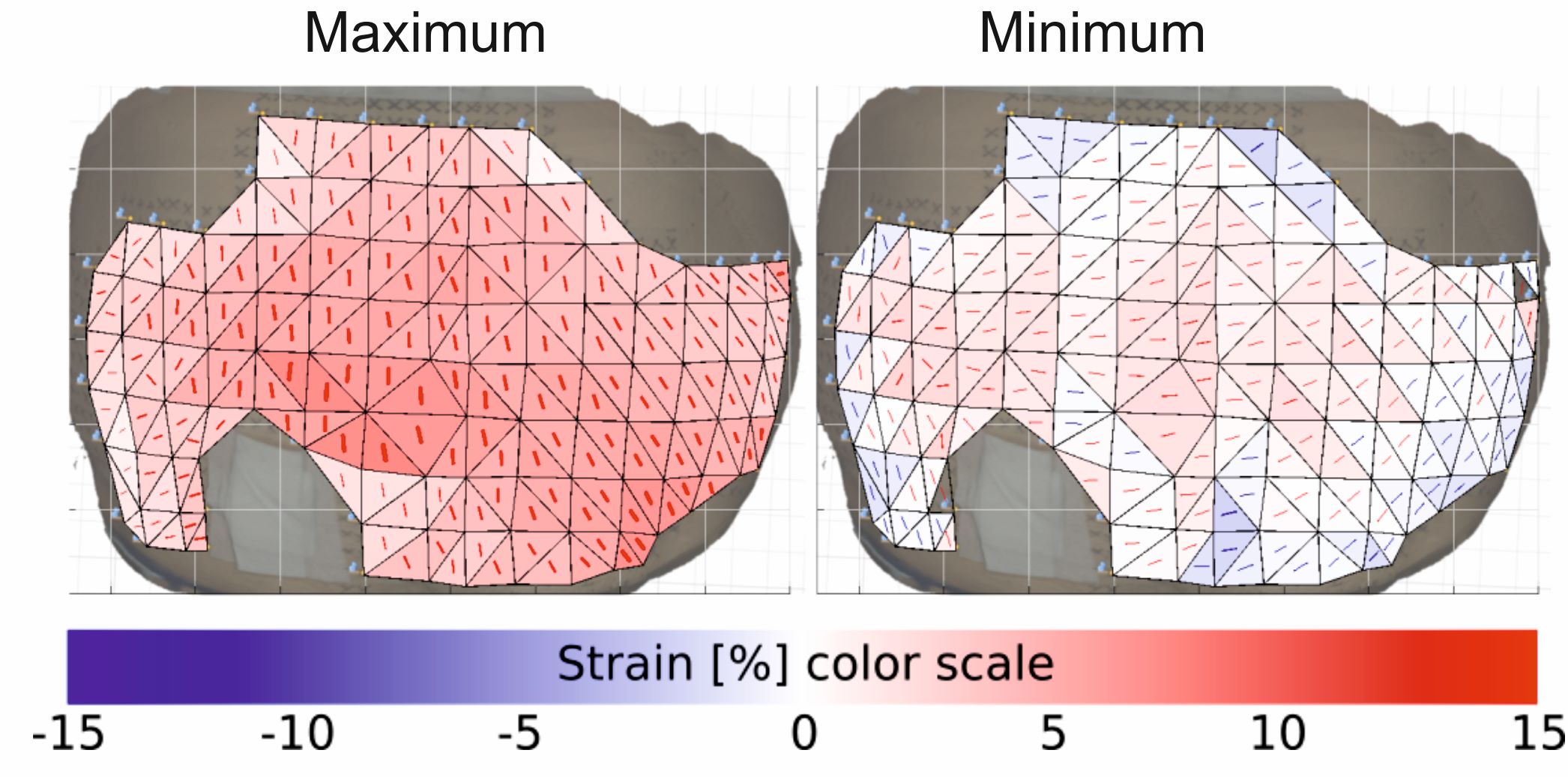}
    \caption{Principal maximum and minimum Cauchy strains with principal directions for Patient P5}
    \label{fig:maps_sub09}
\end{figure}

\begin{figure}[tbh]\centering
    \includegraphics[width=0.8\textwidth]{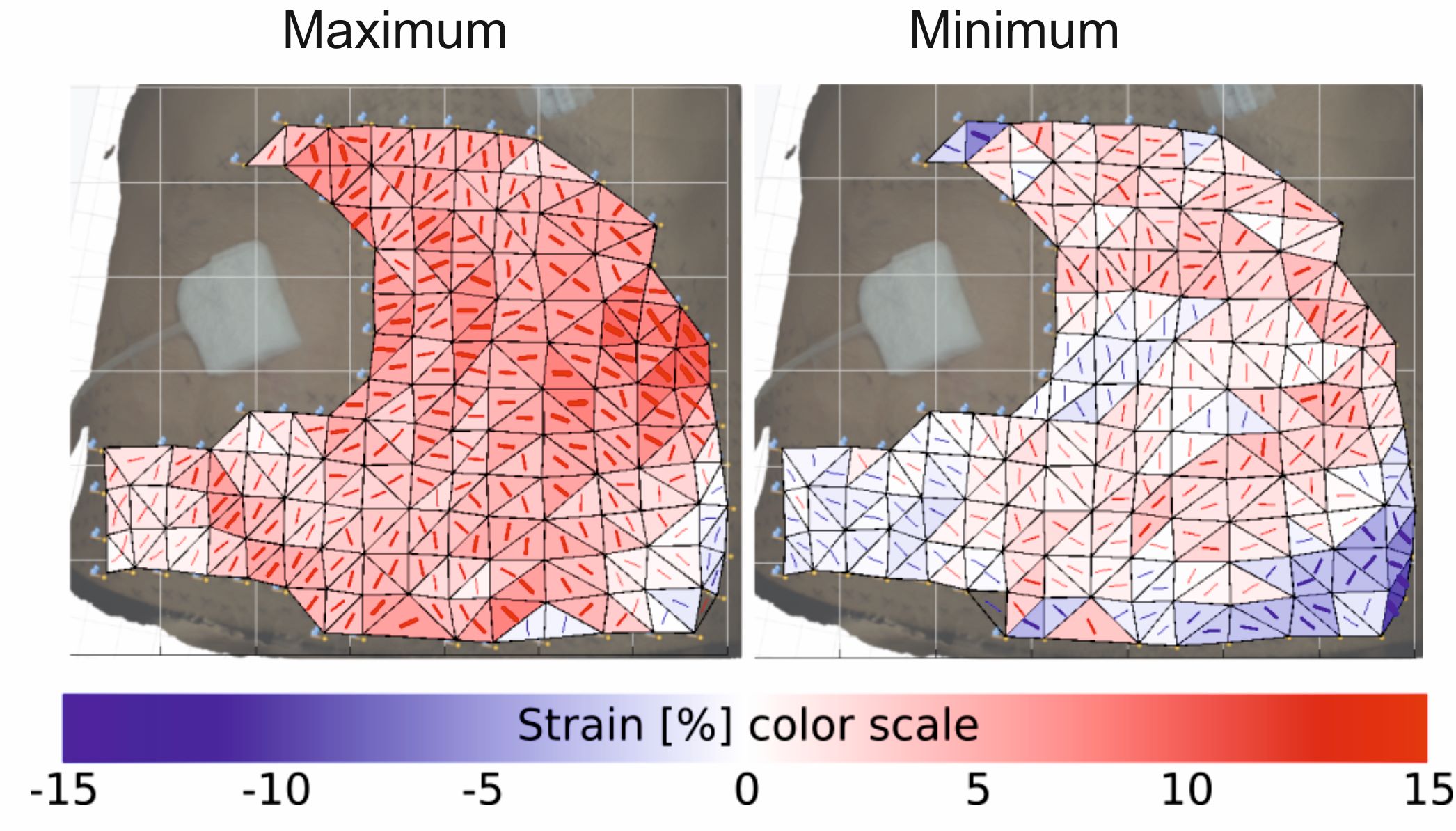}
    \caption{Principal maximum and minimum Cauchy strains with principal directions for Patient P6}
    \label{fig:maps_sub10}
\end{figure}

\begin{figure}[tbh]\centering
    \includegraphics[width=0.8\textwidth]{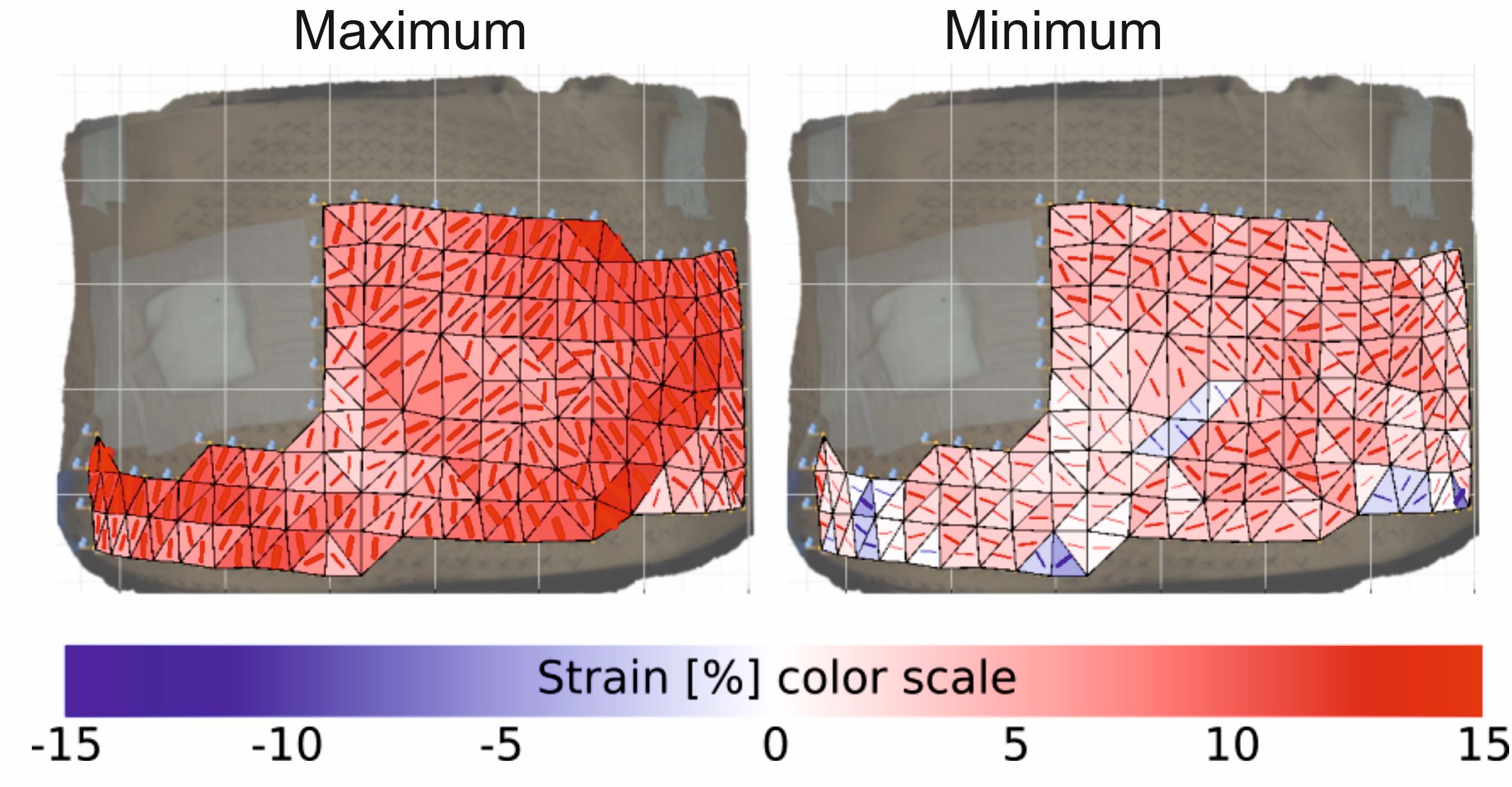}
    \caption{Principal maximum and minimum Cauchy  strains with principal directions for Patient P7}
    \label{maps_sub11}
\end{figure}

\begin{figure}[ht!]
\begin{subfigure}[b]{0.4\textwidth}  
\includegraphics[width=\textwidth]{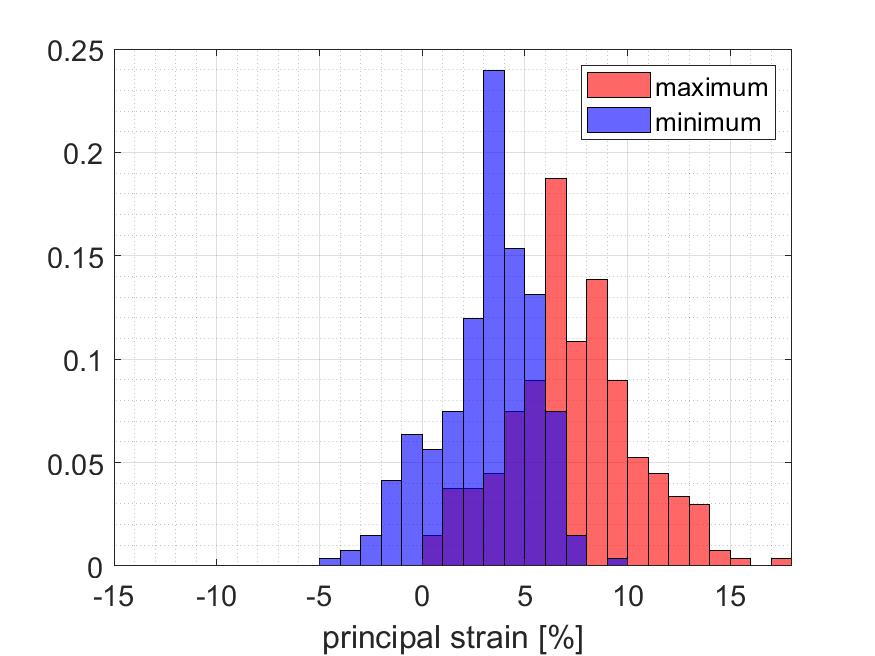}
\caption{Patient P1}
\end{subfigure}
\begin{subfigure}[b]{0.4\textwidth}  
\includegraphics[width=\textwidth]{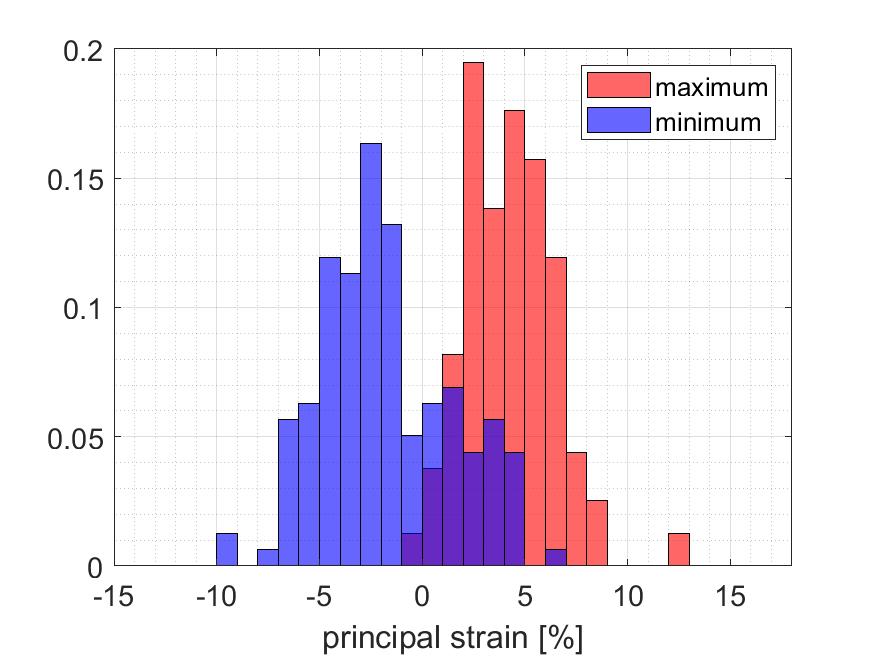}
\caption{Patient P2}
\end{subfigure}

\begin{subfigure}[b]{0.4\textwidth}  
\includegraphics[width=\textwidth]{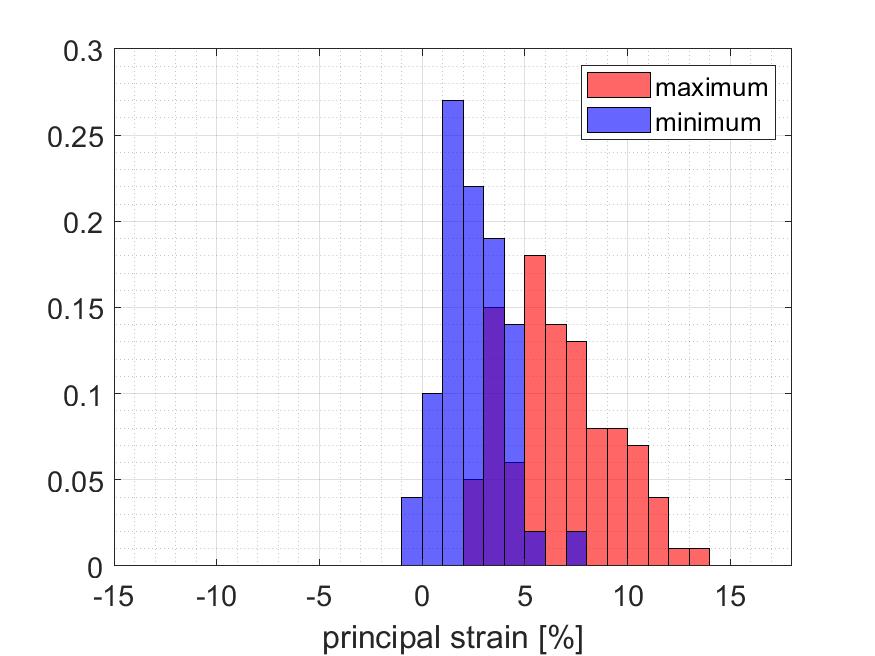}
\caption{Patient P3}
\end{subfigure}
\begin{subfigure}[b]{0.4\textwidth}  
\includegraphics[width=\textwidth]{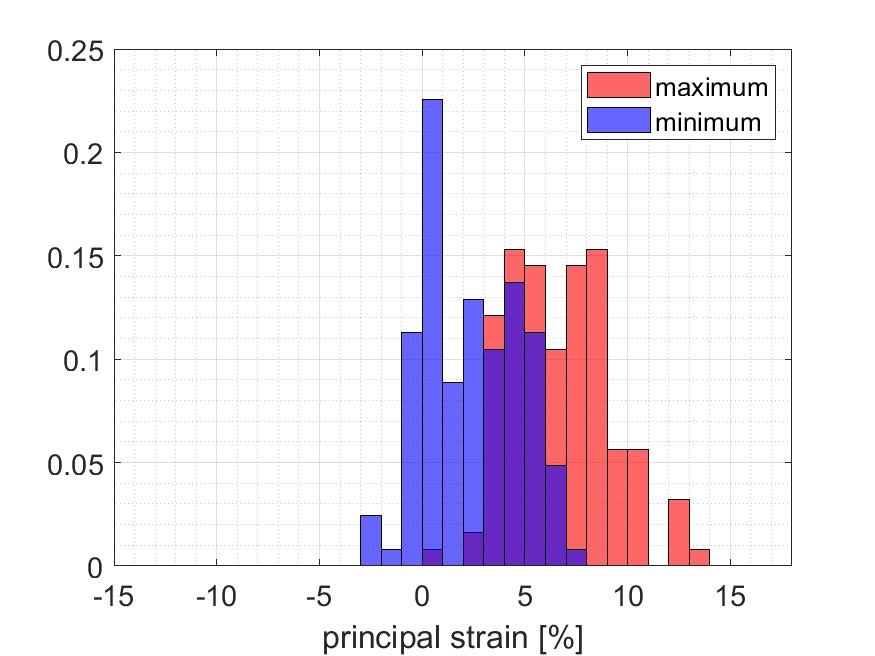}
\caption{Patient P4}
\end{subfigure}

\begin{subfigure}[b]{0.4\textwidth}  
\includegraphics[width=\textwidth]{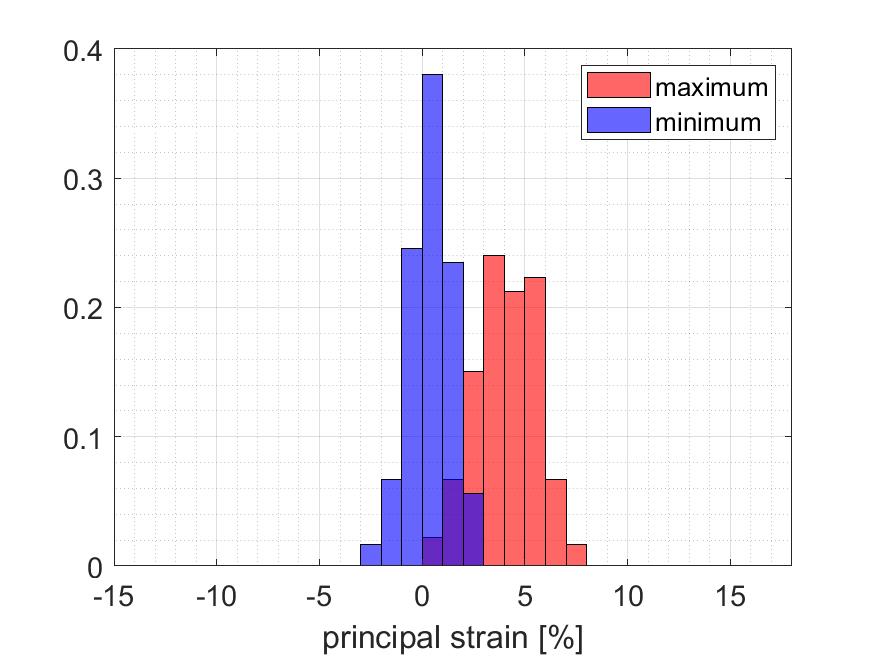}
\caption{Patient P5}
\end{subfigure}
\begin{subfigure}[b]{0.4\textwidth}  
\includegraphics[width=\textwidth]{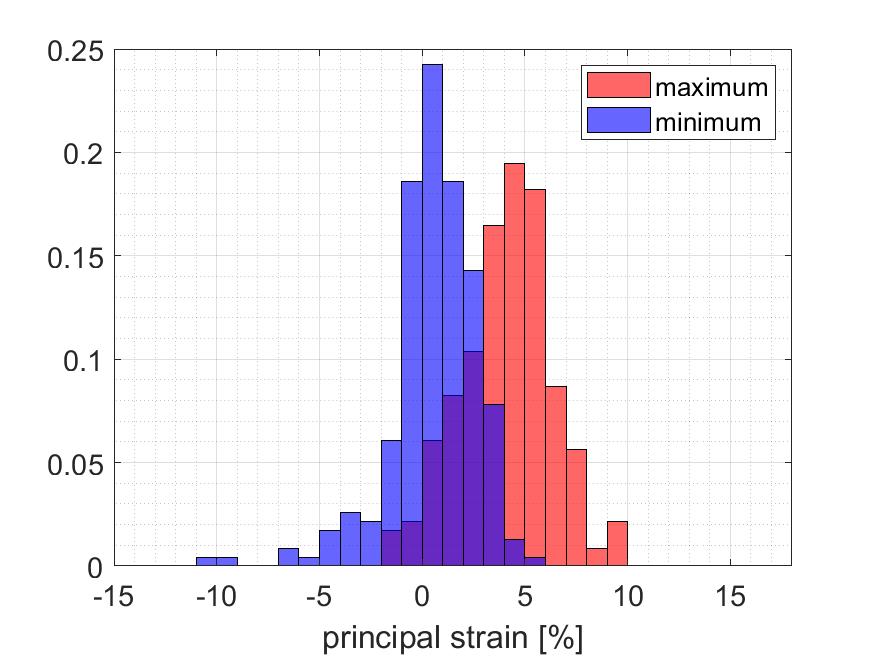}
\caption{Patient P6}
\end{subfigure}

\begin{subfigure}[b]{0.40\textwidth}  
\includegraphics[width=\textwidth]{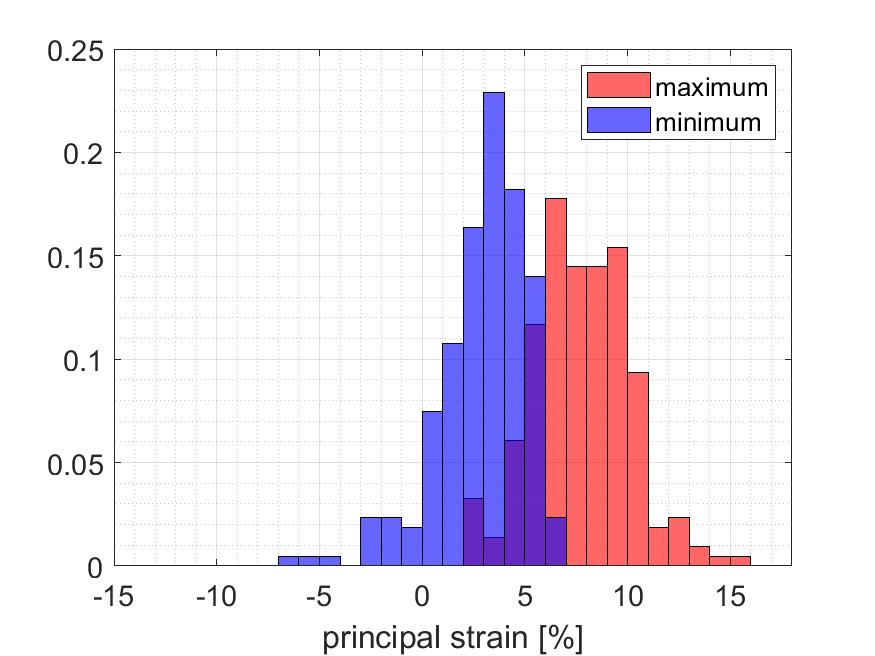}
\caption{Patient P7}
\end{subfigure}

\caption{Normalized histograms of principal Cauchy  strains; maximum $\varepsilon_1$ (red) and minimum $\varepsilon_2$ (blue)} \label{Fig_histograms}
\end{figure}

 Figures  \ref{fig:sub05}--\ref{maps_sub11} show maps of maximum $\varepsilon_1$ and minimum $\varepsilon_2$, the principal strains obtained for tested subjects. The colours in maps are related to the value of principal strains. The principal strain direction  are line-marked in every triangular element. The maps show diverse  spatial distribution of principal strains in the abdominal wall of the patients. The non-uniform character of principal strains is also observable in histograms (Figure \ref{Fig_histograms}) and boxplots (Figure \ref{boxplot}). 
 
Taking into the account the prior belief, the posterior  mean of maximum  $\varepsilon_1$ is a Gaussian variable with mean 12.71\% and standard deviation 1.21\%. That leads to predictive distribution of maximum $\varepsilon_1$ with the same mean and standard deviation 3.52\%. Its coefficient of variation equals 28\% belonging to the variability range observed in biological materials \citep{cook2014biological}. Such variability should be further included in the modelling.  The obtained probabilistic form could be next propagated in  surgical mesh models to consider uncertainty of maximum strains in the analysis of implants \citep{szepietowska2018sensitivity}.

\begin{table}[ht]
\caption{Intraabdominal pressure  and corresponding strain observed in case of each patient}\label{Table_pressure}
\centering
\begin{tabular}{ccccp{0.05cm}ccp{0.05cm}cccc}
\hline
No. &   pressure & \multicolumn{10}{c}{Principal strains* [\%]}\\
\cline{3-12}
 &   [cmH$_2$O]  & \multicolumn{2}{c}{Maximum}& & \multicolumn{2}{c}{Minimum}&& \multicolumn{4}{c}{Median} \\
\cline{3-4}\cline{6-7}\cline{9-12}
 &  & $\varepsilon_1$ & $E_1$ & & $\varepsilon_2$&  $E_2$ && $\varepsilon_1$  &$E_1$ & $\varepsilon_2$ & $E_2$\\

\hline
P1 &15 &	17.0& 18.5 &&	-4.6&-4.4 &&	7.1&7.3 &	3.5 &3.6\\
P2&11.5&	12.4& 13.2&&	-10.0&-9.4  &&	4.1&4.3  &	-2.2 &-2.2\\
P3&11.5&		13.3&14.3 & &	-0.8& -0.7 &&	6.5	&6.7  &2.4 &2.5\\
P4&11&		13.2& 14.1&&		-2.6&-2.5  &&		6.6&6.9 &		2.2 &2.2	\\
P5&18.5&  7.3	& 7.6&&	-2.5&-2.4  &&		4.1	&4.2  &	0.3  &0.3\\
P6&11.5&		10.0&10.5 &&		-10.1&-9.5  &&		4.3&4.4 &		0.7 &0.7	\\
P7&15.5	&	15.6&16.9 &	&	-6.9&-6.6 &&		7.7& 8.0 &		3.4 &3.4	\\
\hline
\end{tabular}

\footnotesize{ * $\varepsilon$ denotes Cauchy strain, $E$ denotes Green-Lagrange strain}
\end{table}

\FloatBarrier

\FloatBarrier

\FloatBarrier

\FloatBarrier

\FloatBarrier

\FloatBarrier

\FloatBarrier
The lowest maximum  $\varepsilon_1$ was observed in the  Patient P5 case, with the highest intraabdominal pressure referring to the same amount of fluid as in other patients (see Table \ref{Table_pressure}). This  implies a high stiffness of the abdominal wall.  The ranges of strains, $\varepsilon_1$ and $\varepsilon_2$, are the narrowest amongst all patients of this case (see Figure \ref{boxplot}). 

The  variability can be seen among all tested patients (see Figures \ref{Fig_histograms} and  \ref{boxplot}). However, similarities between selected  patients may be observed.
The median values are similar between patients of similar age and intraabdominal pressure value, i. e. between Patients P3 and P4 (65 years old, 11.5 cmH$_2$0 and 64 years old, 11.5 cmH$_2$0, respectively), between Patients P5 and P6 (both are 34 years old, but  different intraabdominal pressure: 18.5 and 11.5 cmH$_2$0, respectively) and finally between Patients P1 and P7 (46 years old, 15 cmH$_2$0 and 47 years old, 15.5  cmH$_2$0, respectively). A sole Patient P2 shows a negative median of $\varepsilon_2$. 
In this case negative  values of the $\varepsilon_2$ appear in large part of abdominal surface (Figure \ref{fig:maps_sub06}).

In this study, it can be seen that the values of strains of the outer surface of the abdominal wall are mostly positive. However,   negative strains are also observed (see Figures \ref{fig:sub05}-\ref{fig:maps_sub10},  \ref{Fig_histograms}). 
The presence of negative strains may be linked with  heterogeneity of the abdominal wall related to its complex architecture. Secondly, it can be influenced  by prestrain  observed in living tissues as well as boundary conditions, related to the attachment of tissues to the rib cage and pelvis, as the negative values appear mainly in the vicinity of the chest and hips. In addition, it can also be influenced by the coupling of membrane forces and bending.

The abdominal wall health conditions of patients undergoing measurements is diverse. Patients P2-P5 have or are suspected to have abdominal hernia.  Although, the presence of hernia is expected to affect the results, any hernia effect is observed in this study. The strain range of  Patient P5 is in range 7.3\%--0.5\%, -2.5--2.6\%,  in case of $\varepsilon_1$ and $\varepsilon_2$,   and first principal directions is close to cranio-caudal axis without a visible hernia effect. More patients should be investigated in the light of hernia impact  on the registered strains of abdominal wall. 

The obtained   principal strain directions do not follow the same pattern in all patients, instead they vary along the abdominal wall of individual patient. The Patients P2, P3 and P5 show it more regular. The  first principal strain direction around a mid-line is most clearly aligned along  the cranio-caudal direction  in the case of Patient P5 only. This orientation is consistent with the results of studies on the abdominal walls of human cadavers subjected to intraabdominal pressure described by \cite{LERUYET2020103683}. The disturbance in  principal directions observed in other cases analysed here may be caused by active work of muscles associated with breathing. The presented deformation is mainly related to passive behaviour of myofascial system of abdominal wall under  dialysis fluid introduction. However, some active muscle contribution  of a breathing patient may possibly affect the strain results. 

The change of principal directions is observable in areas close to the peritoneal catheter dressing in  some patients, e.g. P1, P4, P6. The patch  creates a local disturbance in the strain field of the external surface of the abdominal wall. 
Thus, with extreme caution, the directions of principal strains in the area of the patch should be interpreted. Therefore, in  future identification of mechanical properties of abdominal wall based on this study, the part of abdominal wall close to catheter  should be neglected, while  affected by the patch. This is a limitation in capturing the asymmetry of  the abdominal wall.  This asymmetry is indicated in several studies, e.g.,  \cite{jourdan2020abdominal} reported  geometric asymmetry of abdominal muscles. \cite{todros20193d} observed mostly symmetrical displacement of rectus muscle and a more pronounced asymmetry in the case of lateral muscles during muscle contraction. Nevertheless, in the existing research on numerical models of the abdominal wall,  the material parameters are usually  symmetric \citep{he2020numerical}, even a full symmetric model is assumed  with respect to the sagittal plane \citep{pachera2016numerical}.

In addition to patient variability in factors such as age, BMI, and health condition, the patients differ in the  measured area. This is bounded by a patch covering the catheter placed in different regions of the abdominal wall.

\begin{figure}[tbh]\centering
\includegraphics[width=\textwidth]{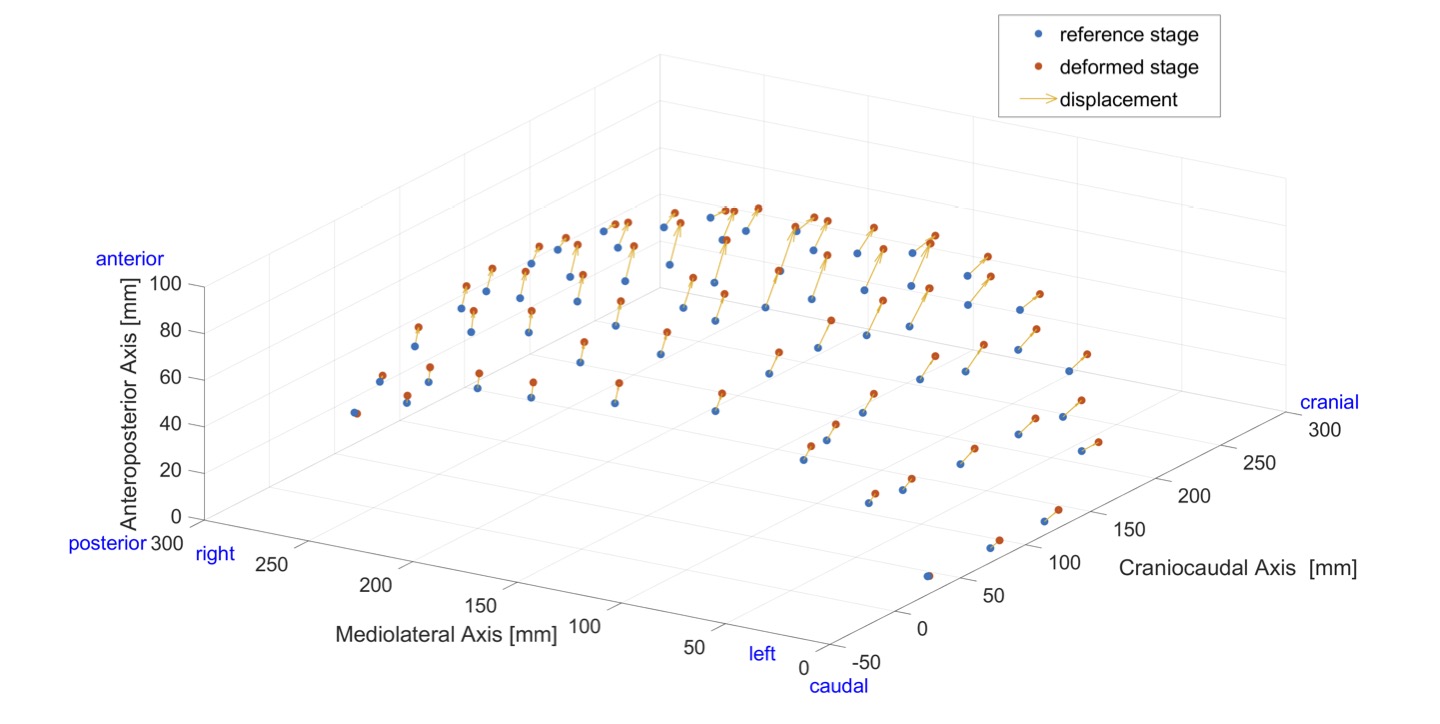}
\caption{Patient P3 displacements (range from 0.6 to 19.3    mm)}
 \label{Fig_disp_p3}
\end{figure}

\begin{figure}[tbh]\centering
\includegraphics[width=\textwidth]{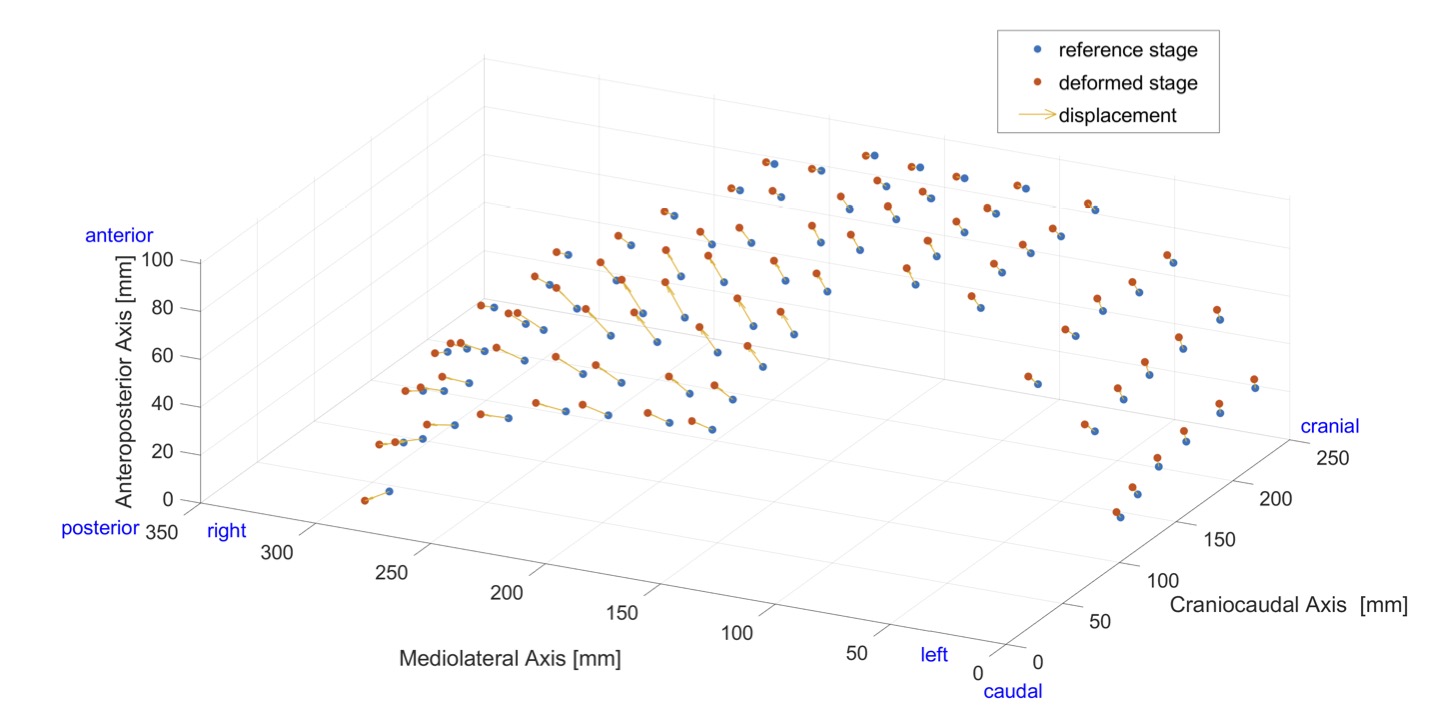}
\caption{Patient P4 displacements (range from 3.3 to 19.9   mm)}
 \label{Fig_disp_p4}
\end{figure}

\begin{figure}[tbh]\centering
\includegraphics[width=\textwidth]{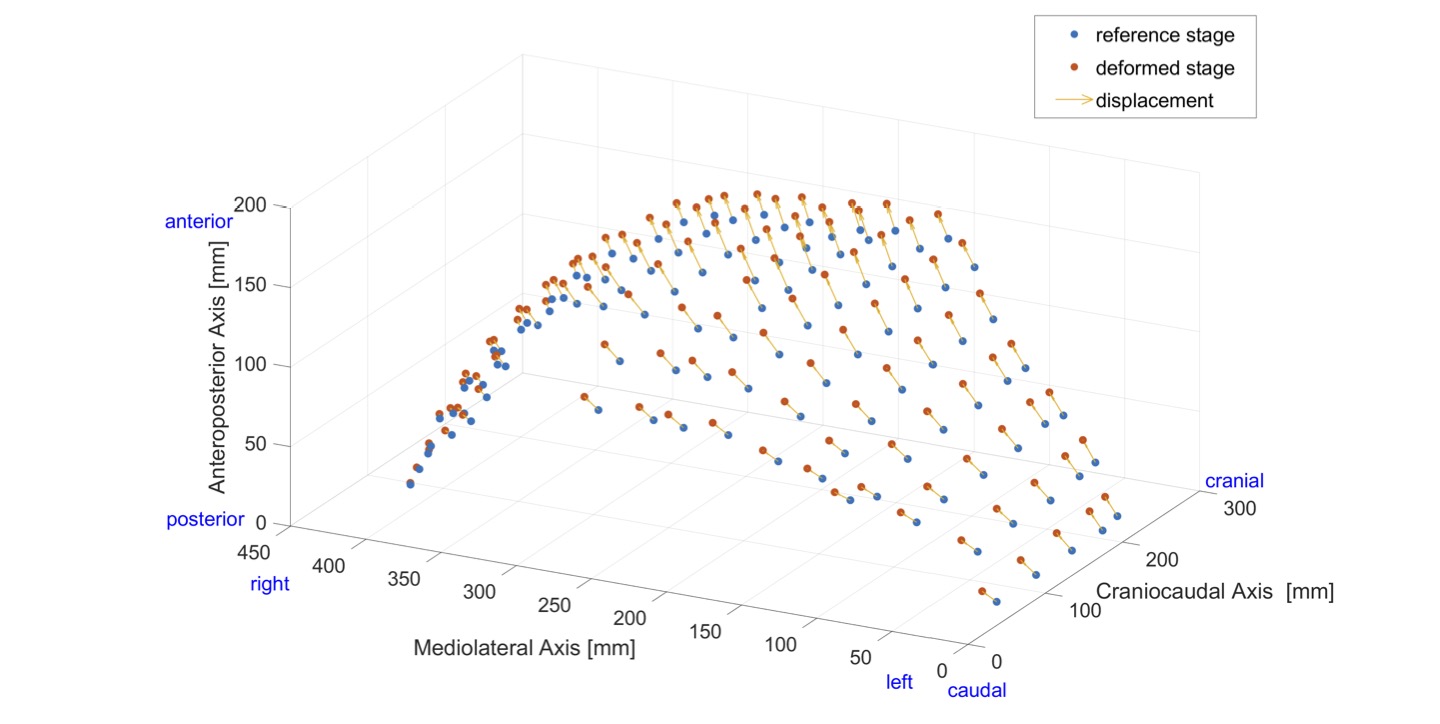}
\caption{Patient P5 displacements (range from 4.0 to 21.5  mm)}
 \label{Fig_disp_p5}
\end{figure}

\begin{figure}[tbh]\centering
\includegraphics[width=\textwidth]{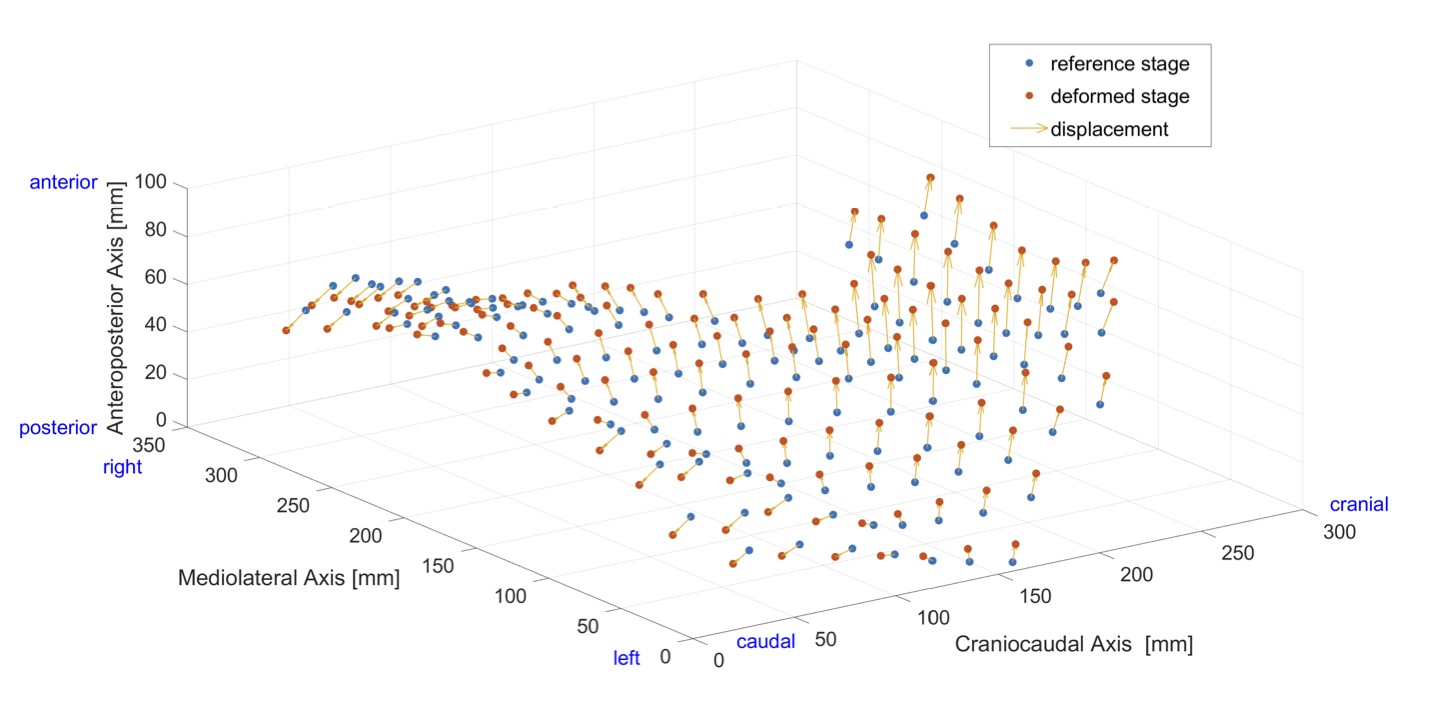}
\caption{Patient P6 displacements (range from 6.2 to 22.4 mm)}
 \label{Fig_disp_p6}
\end{figure}

Figures \ref{Fig_disp_p3}--\ref{Fig_disp_p6}  show the abdominal wall nodal displacements of  two oldest and two youngest patients, respectively, which includes also patients with the highest BMI (Figure \ref{Fig_disp_p3}) and the lowest BMI (Figure \ref{Fig_disp_p6}). The maximum displacement in cases of the oldest patients, P3 and P4 (similar age but quite different BMI), is similar and equal to 19.3 mm and 19.9 mm respectively (see Figures \ref{Fig_disp_p3} and \ref{Fig_disp_p4}). In the case of the youngest patients, P5 and P6, both at age of 34, and again different BMI, the registered maximum displacement of the abdomen is higher (21.5 mm and 22.4 mm) but again these values are quite similar to each other (see Figures \ref{Fig_disp_p5} and \ref{Fig_disp_p6}). The plot of displacements for P6 differs from other patients. This patient has the lowest BMI (20) that determines the shape of the abdominal wall and probably also its mechanical behaviour.

Although other studies presented principal strains of abdominal wall under diverse conditions (\textit{in}  or \textit{ex vivo}, loading condition etc.),   the present results  are partially referred to the values reported in literature. 
\begin{itemize}
    \item our past study  \citep{szymczak2012investigation} adressed strains on the living abdominal wall under activities like bending or stretching were analysed. That study also shows a high inter patients variability. The observed  range of strains is higher there, in a given oblique direction the mean values of strains reaches even 34\%.  However, the investigated group was younger (23--25 years old) and healthy. Nevertheless, a high difference in strain ranges in this study and in the present research, was  mainly caused by different loading conditions, inducing active muscle behaviour, while the present study refers to their mostly passive behaviour under  slow change of intraabdominal pressure.  Furthermore, torso bending is considered decisive to trigger high deformations of he abdominal wall. This was shown in the \textit{in vivo} study by \cite{lubowiecka2020vivo}, here the elongation of surgical meshes implanted to  the abdomen was tested while under torso bending. A numerical study by \cite{szymczak2017two} suggests that deformation of abdominal wall caused by such activities may provoke higher forces in the joints connecting surgical mesh to the abdominal wall than the loading produced by intraabdominal pressure.
    \item \cite{LERUYET2020103683} presented  strains in myofascial abdominal walls (without fat and subcutaneous fat)  of human cadavers subjected to cycles of pressure, reporting high variability amongst studied samples. The research was aimed at comparing suturing techniques,  the results for intact \textit{linea alba} were also shown. The maximum strains and their directions of this inquiry are similar to the present study when refer to a similar range of pressure applied to the abdominal wall. A single sample  reveals larger strain value in the sample domain (Green-Lagrange strains close to 20\% in the point in the midline).
    
    \item In the \textit{in vivo} research on protruded and contracted human abdominal walls described by \cite{breier2017evaluation}, the maximum observed strain reached 60\%. Similarly to our study, the authors observed  high variability among the patients.  
    \item \cite{podwojewski2013mechanical} obtained, in their \textit{in vitro} study, an average value of 13.7\% of the first principal Lagrange strain on the outer surface of the abdominal wall of a porcine abdominal wall subjected to  pressure. The pressure was equal to 50 mmHg (around 68 cmH$_2$0),  higher compared to the present research.

\end{itemize}

It can be stated that the range of principal strains of the ongoing research is consistent with the results reported in the literature related to strain in the abdominal wall subjected to similar loading.  

It should be noted that the abdominal wall is a complex multi-layered structure of components with different fibers alignments. 
It contains the: skin, subcutaneous tissue, superficial fascia, rectus abdominis muscle,  external oblique muscle, internal oblique muscle, transversus abdominis muscle, transversalis fascia, preperitoneal adipose and areolar tissue, and peritoneum. Nerves, blood vessels, and lymphatics are present throughout. The contour and thickness of the abdomen is dependent upon age, muscle mass, muscle tone, obesity, intra-abdominal pathology, parity, and posture. Integrity of the anterior abdominal wall is primarily dependent upon the abdominal muscles and their conjoined tendons. The average thickness of rectus abdominis muscle and abdominal subcutaneous fat tissue measured by \cite{kim2012thickness} with the use of computer tomography, was 10 mm, and the average subcutaneous tissue thickness equalled 24 mm.
Nevertheless, in our study only the  strains of the outer surface of the abdominal wall are investigated. Therefore the connective tissues and muscles considered important are not directly observed. The latter  issue is a possible limitation of the study.  However
\cite{tran2014contribution} showed that the contribution of skin and adipose tissue in the strains of the abdominal wall under pressure is not as decisive as, say, rectus sheath.

Next limitation is that  nonlinear behaviour of the abdominal wall in response to intra-abdominal pressure is not captured in this study. Our aim was to interfere with the medical procedure as little as possible.  Therefore, since standard methodology of peritoneal dialysis does not assume measurements of IAP in intermediate states, the pressure measurements have been performed  after filling  peritoneal cavity with 2000 ml of dialysis fluid.The measurements need to be limited for the sake of patient’s safety because additional measurements of intra-abdominal pressure would have been associated with an increased risk of peritoneal infection.

This study is focused on strains. However,  the results can be further used in the identification of the mechanical properties of abdominal wall. That will require deep consideration of various  aspects of this issue, such as  nonlinearity, anisotropy, etc.  In addition, residual stresses can affect the mechanical response of the entire system, which should be considered in future simulations and the identification of abdominal stiffness \citep{rausch2013effect}. The reference configuration of drained abdominal wall analysed in the current study may not be stress free.

It should be noted that intraabdominal pressure is considered  a load in much of the research on numerical abdominal wall modelling \citep{pachera2016numerical,hernandez2013understanding}. The presented results   can be incorporated to validate numerical models of the anterior abdominal wall. 

\section{Conclusions}

The methodology proposed in the study brings the deformation  of the living human abdominal wall corresponding to the intraabdominal pressure measured during peritoneal dialysis. This approach is successfully applied for a  group of patients with various abdominal wall health status.  The dedicated experimental stand simplifies geometry measurement of a living object. Although, the post-processing of the photogrammetric data is more time consuming when compared to commercial digital image correlation systems, the measurement alone can be performed relatively quickly. This fact, combined with the ease  and quick assembly, the absence of wires and the relatively small room required, makes the mobile experimental stand suitable for use in hospitals during standard PD fluid exchange.

The study presents the values and spatial distribution of  strains observed on external surface of abdominal wall of each patient. Cauchy and Green-Lagrange strains are shown.  The maximum principal Cauchy strains vary from 7.3 to 17\%,  the median of first principal directions varies from 4.1 to 7.7\% throughout the patients. The set of maximum values shows a mean 12.7\%. The observed intraabdominal pressure ranges from  11 to 18.5 cmH$_2$O. The obtained principal strains  and their directions indicate  variability in a patient domain. 

High variability observed  between the subjects in each study indicates high mechanical parameter variability of the abdominal wall. This justifies a need for patient-specific approach towards treatment optimisation   of ventral hernias. However, in the future, a larger group of patients should be tested to investigate the effect of various parameters,  e.g. the age, BMI or health condition, to possibly affect the  range of strains of human living abdominal wall.

The presented work is the next step towards the recognition of  mechanics of  living  human  abdominal wall. The presented approach can be further expanded to identify the mechanical properties  of human abdominal wall based on \textit{in vivo} measurements, performed in non-invasive and relatively inexpensive way. The results can also be  applied to validate other computational models of the abdominal wall. The research entirety is aimed at improving abdominal hernia repair as reliable computational models will enable \textit{in silico} analysis and optimisation of abdominal hernia treatment parameters.

\section*{Acknowledgements}

We would like to thank the staff of Peritoneal Dialysis Unit Department of Nephrology Transplantology and Internal Medicine Medical University of Gda\'nsk and Fresenius Nephrocare (dr Piotr Jagodzi\'nski, nurses Ms Gra\.zyna Szyszka and Ms Ewa Malek) for their help in accessing the patients, performing PD exchanges and measurements of the IPP.

This work was supported by the National Science Centre (Poland) [grant No. UMO-2017/27/B/ST8/02518]. Calculations were carried out partially at the Academic Computer Centre in Gda\'nsk.


\bibliography{07_mybibfile}

\end{document}